\documentclass[preprint]{aastex62}

\usepackage{amsmath,bm,mathrsfs,graphicx}
\usepackage{epstopdf}

\shorttitle{2D PCA de-fringing}
\shortauthors{Casini \& Li}

\begin{document}

\title{Removal of Spectro-Polarimetric Fringes by 2D PCA}

\correspondingauthor{R. Casini}
\email{casini@ucar.edu}

\author{R. Casini}
\author{W. Li}

\affiliation{High Altitude Observatory, National Center for Atmospheric Research$^\dagger$\footnote[0]{$^\dagger$The National Center for Atmospheric Research is sponsored by the National Science Foundation.} \\
P.~O.~Box 3000, Boulder, CO 80307-3000, USA}



\begin{abstract}
We investigate the application of 2-dimensional Principal Component Analysis (2D PCA) to the problem of removal of polarization fringes from spectro-polarimetric data sets. We show how the transformation of the PCA basis through a series of carefully chosen rotations allows to confine polarization fringes (and other stationary instrumental effects) to a reduced set of basis ``vectors'', which at the same time are largely devoid of the spectral signal from the observed target. 
It is possible to devise algorithms for the determination of the optimal series of rotations of the PCA basis, thus opening the possibility of automating the procedure of de-fringing of spectro-polarimetric data sets. We compare the performance of the proposed method with the more traditional Fourier filtering of Stokes spectra.
\end{abstract}



\section{Introduction} \label{sec:intro}

The advent of new instrumentation for the investigation of solar phenomena---in particular, with regard to the short- and long-term variability of the Sun's magnetic activity, and its impact on the near-Earth environment---has put unprecedented demands on the conception of data management plans and infrastructures, which aim to deliver reliable, science-ready data products to the community in a timely fashion.

The U.S.\ community is getting ready for the completion and first light of the 4-m Daniel K.\ Inouye Solar Telescope \cite[DKIST;][]{Tr16} in early 2020, with its suite of complex post-focus instruments, most of which will have polarimetric capabilities, in order to detect the subtle signatures of the Sun's magnetic field in its light spectrum. The 1.6-m Goode Solar Telescope \cite[GST;][]{Ca10} has already started to reveal the complexity of the solar spectrum when unprecedented higher spatial and temporal resolutions are attained in solar observations. Other countries are pursuing similar endeavors, such as India's 2-m National Large Solar Telescope \cite[NLST;][]{Ha10}, the 4-m European Solar Telescope \cite[EST;][]{Co13}, and the 8-m annular-mirror Chinese Giant Solar Telescope \cite[CGST;][]{Li14}.

Beyond the intrinsic complexity of the theoretical problem of how polarized radiation is formed and transported through the solar atmosphere \cite[e.g.,][for a description of the various mechanisms at play]{St94,LL04,CL08,TB10}, the need to detect and confidently measure very small levels of polarization (down to 0.01\% of the intensity, for the most demanding scientific applications) puts very hard requirements on the identification and removal of instrumental artifacts from the detected signals.

One recurrent issue in spectro-polarimetric instruments is the appearance of polarization ``fringes'' that overlap with the actual spectral signal from the 
observed target. These fringes are interference patterns that are produced by the
presence of optical elements in the system of the telescope and the instrument
having varying phase retardance properties (e.g., around their optical axis). Such components include polarization modulators, polarizing beam-splitters, and any 
optical element where parallel optical interfaces may occur (e.g., interference filters, detector windows). These fringes have the appearance of more or less
regular 2-dimensional (2D) patterns, preferentially arranged along the spectral dimension of the data (see Figure 1). We refer to review studies of polarized fringes \citep{Li91,Se03,Cl04} for a thorough description of this phenomenon, and to recent work by \cite{Ha17} on the use of Berreman calculus for the modeling of fringes in polarimetric instrumentation.

The most common fringe correction methods used in spectro-polarimetric data reduction are Fourier filtering and derivations from it, such as wavelet analysis \citep{Ro06}. However, Fourier filtering only works satisfactorily when the frequency domains of the fringes and of the target spectral signal (in the spectral and/or spatial domains) are clearly separated, and it is severely limited when the fringe pattern's period or amplitude vary over the image. Wavelet analysis is a powerful method for removing smoothly varying fringes in flat-field images. However, it becomes difficult to control in the presence of strong spectral signals from the observed target.

\begin{figure}[t!]
\centering
\includegraphics[width=5in]{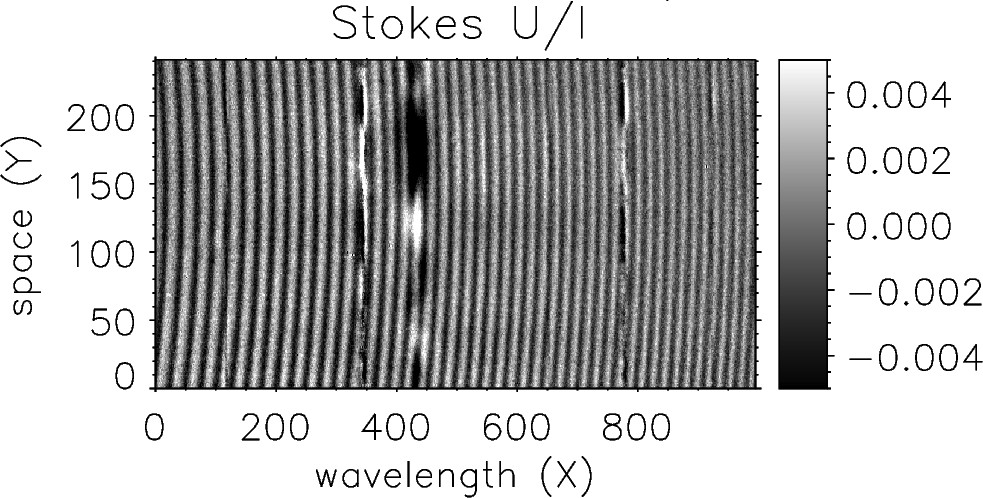}
\caption{\label{fig:example}
Example of a polarized spectrum (in this case, the fractional linear polarization, Stokes $U/I$) of the solar radiation around the 1085\,nm. The three dominant features from the left are: the photospheric line of \ion{Si}{1} at 1082.7\,nm ($X\simeq 350$); the chromospheric triplet of \ion{He}{1} around 1083\,nm ($X\simeq 430$); and the photospheric line of \ion{Si}{1} at 1084.4\,nm ($X\simeq 780$). We note the presence of strong polarization fringes across the entire spectrum.}
\end{figure}

In this paper, we extend previous work on the identification and isolation of polarization fringes by Principal Component Analysis \cite[PCA;][]{Pe01,Jo02}.
\cite{Ca12} considered the implementation and performance of a 2D PCA algorithm based on the method described by \citeauthor{Ya04} (\citeyear{Ya04}; hereafter, Y-PCA). In that approach, the spectral data gets ``contracted' over the spatial dimension before performing PCA decomposition. As a result, the PCA basis of \emph{eigenfeatures} consists of spectro-polarimetric profiles (rather than 2D images similar to Figure~\ref{fig:example}), which are akin to spatial averages of the principal components (PCs) of the data.
One advantage of the Y-PCA approach is the fast convergence of the singular-value series, which in theory allows to limit the number of PCs used for the data reconstruction to a small set of the lowest order basis vectors. A beneficial side effect is that unwanted, high-order spatial modulations of the signals that tend to cancel out when spatially averaged, such as detector noise, get confined to the lower end of the ordered set of PCs, so a low-pass cut of this set is a very efficient way to remove these unwanted features from the data \cite[see][]{Ca12}.
Not surprisingly, these benefits come at a cost. It may often be hard to tell fringes and spectral signals apart (with the exception of the most dominant fringes), and the compression of spatial information in the Y-PCA algorithm causes high-order spatial variations of the spectral signals to be distributed over many PCs, which therefore must be included for the reconstruction of the signal, if such spatial variations are to be preserved, e.g., in high resolution observations.

In this work, we consider the application to the polarization fringe problem of the original 2D PCA algorithm of \citeauthor{TP91} (\citeyear{TP91}; hereafter, TP-PCA), which has traditionally been applied to the problem of face recognition. The possibility of using this approach to fringe removal was briefly touched upon in the conclusions of \cite{Ca12}. One downside of the algorithm pointed out in that paper is that the singular value series shows a much slower convergence than in the Y-PCA method, and the dimensionality of the basis of eigenfeatures corresponds to the number of images in the data set to be analyzed, which can in some cases be a small number (even just one). In practice, this implies that often one must keep the entire set of PCs in order to reconstruct the data without significant loss of information, especially for small sets of images.

On the other hand, the advantage of the TP-PCA method is that the contribution of the fringes to the data is immediately recognizable in the set of eigenfeatures, which in principle should make the removal of such contribution an easier task than in the Y-PCA method. However, since most or all of the PCs must be included in the reconstruction in order to preserve the original spatial and spectral information of the data, no practical benefit can come from this, unless the fringes somehow are confined to just a few eigenfeatures essentially devoid of spectral signal, which then can be dropped from the data reconstruction without any risk of signal loss.

The purpose of this paper is to apply a simple manipulation technique of the PCA basis, in order to optimally attain the isolation of the fringes in a few basis vectors. This technique relies on the intrinsic orthogonality of the PCA basis, and uses rotations on the 2D subspaces generated by all possible pairs of eigenfeatures in the basis, so that the corresponding transformed 2D subspace has one of the two basis vectors clean of fringes (as far as possible). By iteratively processing the entire basis, it is possible to \emph{automatically} confine the fringes to a minimal set of transformed basis vectors. Because this transformation of the PCA basis is constructed as a sequence of isometries, the process naturally preserves the orthonormality of the original PCA basis after rotation.

The idea of rotations applied to the PCA basis in order to bring out particular physical properties of the original data set is not new, and it has also been applied to the problem of spectro-polarimetric inversion of photospheric lines \citep{SL02}. The technique presented here can indifferently be applied to both the TP-PCA and Y-PCA approaches. In the case of the Y-PCA method, this manipulation technique can significantly improve the separation of fringes and spectral signals in the set of basis vectors, compared to the direct application of the method as presented in \cite{Ca12}.

In Section~\ref{sec:theory}, we present the mathematical concepts and assumptions for the separation of spectral signals and unwanted instrument artifacts in a spectro-polarimetric data set. In Section~\ref{sec:examples}, we show some applications of the basis manipulation technique to the PCA of Stokes data for both the TP-PCA and Y-PCA approaches. The rotation algorithm is summarized in App.~\ref{app:B}.

\section{Formulation of the problem} \label{sec:theory}

We consider a 2D spectro-polarimetric data set, consisting of an ensemble of $N$ 
``Stokes images'', having spectral wavelength (or frequency) as the $x$ coordinate, and the spatial position along the slit of a grating-based spectrograph as the $y$ coordinate. Alternatively, the data may come from an imaging spectro-polarimeter, where the spectral signal is acquired through wavelength scanning of a tunable filter instrument (e.g., using a Fabry-Perot interferometer; \citealt{Ca06}). The data can therefore be represented by the set 
$\mathscr{D}\equiv\{\mathscr{D}_i(x,y)\}_{i=1,\ldots,N}$.

The implementation of 2D PCA according to the TP-PCA method allows to represent each image in the data set as a linear expansion
over a basis (i.e., an orthonormal set) of eigenfeatures $\bm{e}_j(x,y)$, where $j=1,\ldots,N$,
\begin{equation} \label{eq:expans.d}
\mathscr{D}_i(x,y)=\sum_{j=1}^N c_{ij}\,\bm{e}_j(x,y)+\bar{\mathscr{D}}(x,y)\;,\qquad i=1,\ldots,N\;,
\end{equation}
where $\bar{\mathscr{D}}(x,y)$ is the average image of the data set. 
The orthonormality condition corresponds to the following definition of inner product on the space $\mathfrak{E}$ generated by the basis $\{\bm{e}_j(x,y)\}_{j=1,\ldots,N}$,
\begin{equation} \label{eq:inner}
\langle \bm{e}_j,\bm{e}_k \rangle
\equiv \int\limits_{D_{XY}}\mathrm{d}x\,\mathrm{d}y\;
\bm{e}_j(x,y)\,\bm{e}_k(x,y)=\delta(j,k)\;,
\end{equation}
where $D_{XY}$ is the area of the image, and $\delta(j,k)$ is Kronecker's $\delta$. In the usual implementation of the TP-PCA algorithm \cite[see][]{TP91}, the basis $\{\bm{e}_j(x,y)\}_{j=1,\ldots,N}$ is determined via diagonalization of the $N\times N$ covariance matrix $\textrm{Cov}(\mathscr{D},\mathscr{D})$. 
This can be calculated through a simple matrix multiplication,
\begin{equation} \label{eq:cov}
\textrm{Cov}(\mathscr{D},\mathscr{D})=(\bm{\mathsf{D}}-\bm{\bar{\mathsf{D}}})^T
(\bm{\mathsf{D}}-\bm{\bar{\mathsf{D}}})\;,
\end{equation}
where $\bm{\mathsf{D}}$ is a $(N_X N_Y)\times N$ matrix, whose $i$-th column corresponds to the $\mathscr{D}_i(x,y)$ array of size $N_X\times N_Y$ reformed into a column vector of length $N_X N_Y$ (see App.~\ref{app:A}), whereas $\bm{\bar{\mathsf{D}}}$ is a similarly constructed matrix, where each column is the average of the set of reformed $\mathscr{D}_i(x,y)$ arrays across the map.\footnote{Equation~(\ref{eq:cov}) corrects eq.~(6) of \cite{Ca12} in their description of the \cite{TP91} method.}
If we indicate with $\bm{\mathsf{U}}$ the (orthogonal) matrix of the column eigenvectors of (\ref{eq:cov}), then the normalized basis of eigenfeatures is given by the set (cf.\ App.~\ref{app:A}, eq.~(\ref{eq:2D_basis.app}))
\begin{equation} \label{eq:2D_basis}
\bm{e}_j(x,y)=\lambda_j^{-1/2} \sum_i \left[\mathscr{D}_i(x,y)-\bar{\mathscr{D}}(x,y)\right] U_{ij}\;,\qquad j=1,\ldots,N\;,
\end{equation}
where the $\lambda_j$'s are the corresponding eigenvalues. It is straightforward to verify that this is indeed an orthonormal set (see App.~\ref{app:A}). 

It is important to remark that, in general, the average image $\bar{\mathscr{D}}(x,y)\notin\mathfrak{E}$, when the basis is determined via diagonalization of the covariance matrix (\ref{eq:cov}). This can also be gleaned directly from eqs.~(\ref{eq:expans.d}) and (\ref{eq:2D_basis}).

In practice, the data will include the desired spectro-polarimetric signal ($\mathscr{S}$) 
of the target science diagnostics of the observations alongside with unwanted instrumental signatures, 
such as detector artifacts and polarization fringes ($\mathscr{F}$), so that $\mathscr{D}_i(x,y)=\mathscr{S}_i(x,y)+\mathscr{F}_i(x,y)$, for $i=1,\ldots,N$. The problem of de-fringing the data obviously corresponds to the removal of the contribution $\mathscr{F}_i(x,y)$ for each image in the data set. Because the average image $\bar{\mathscr{D}}(x,y)$ will in general contain unwanted instrumental signal---unlike in the problem of face recognition to which the TP-PCA algorithm is usually applied---the inability of the basis $\{\bm{e}_j(x,y)\}_{j=1,\ldots,N}$ to generate $\bar{\mathscr{D}}(x,y)$ is highly inconvenient. 
In fact, the need to add back the average image to the reconstructed data, according to the standard algorithm, implies that some of the unwanted instrumental signal is uncontrollably put back into the reconstructed data.
For this reason, we adopt a variant of the algorithm, where the basis is determined instead via diagonalization of the correlation matrix (see App.~\ref{app:A}, eq.(\ref{eq:corr.app}))
\begin{equation} \label{eq:corr}
\textrm{Corr}(\mathscr{D},\mathscr{D})=\bm{\mathsf{D}}^T \bm{\mathsf{D}}\;,
\end{equation}
so that eqs.~(\ref{eq:expans.d}) and (\ref{eq:2D_basis}) become, respectively,
\begin{eqnarray} \label{eq:expans.corr}
\mathscr{D}_i(x,y)&=&\sum_{j=1}^N c_{ij}\,\bm{e}_j(x,y)\;,\qquad i=1,\ldots,N\;, \\
%
%
\label{eq:2D_basis.corr}
\bm{e}_j(x,y)&=&\lambda_j^{-1/2} \sum_i \mathscr{D}_i(x,y)\,U_{ij}\;,\qquad j=1,\ldots,N\;.
\end{eqnarray}
%

\begin{figure}[p!]
\centering
\includegraphics[width=7in]{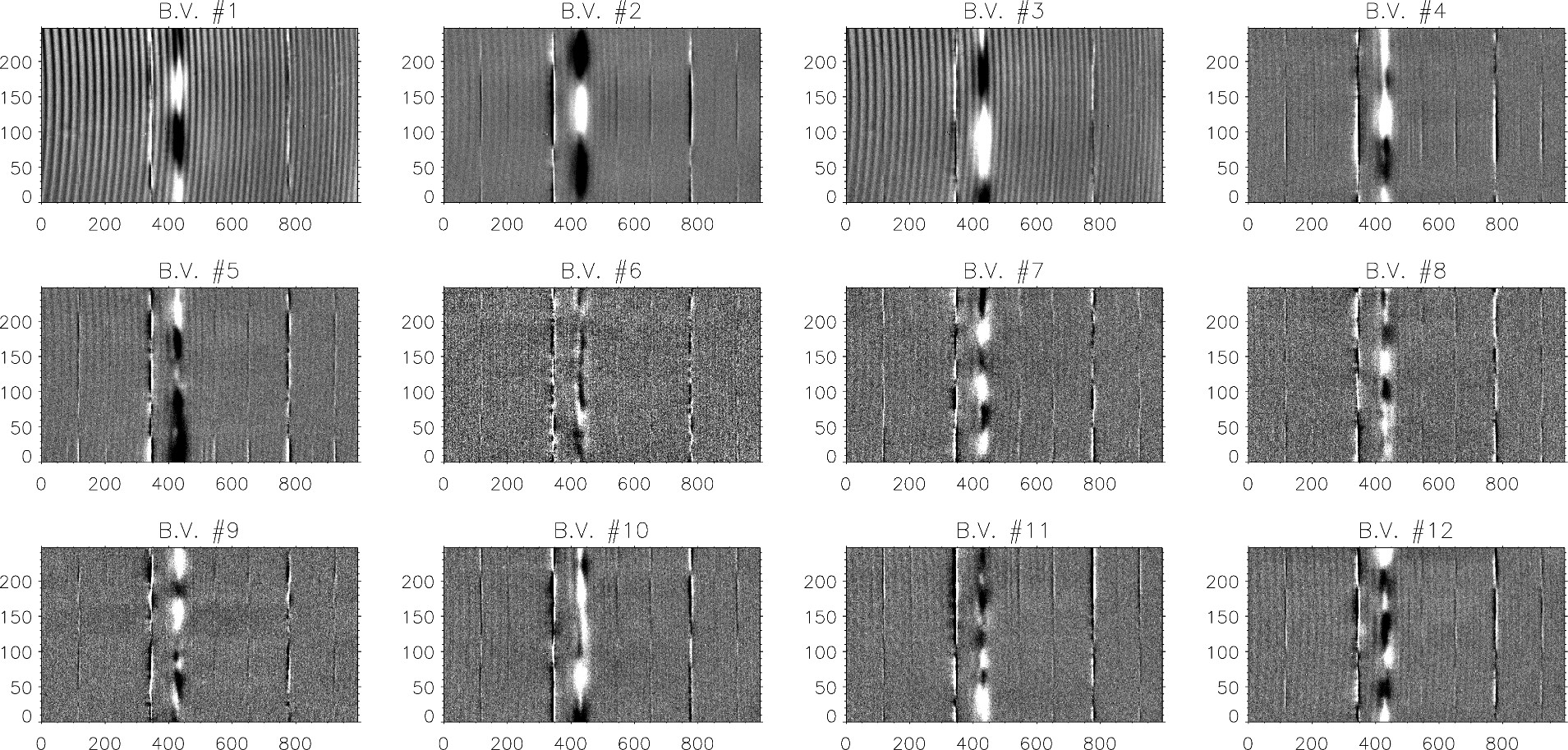}
\vbox{\medskip Original PCA basis\bigskip}
\includegraphics[width=7in]{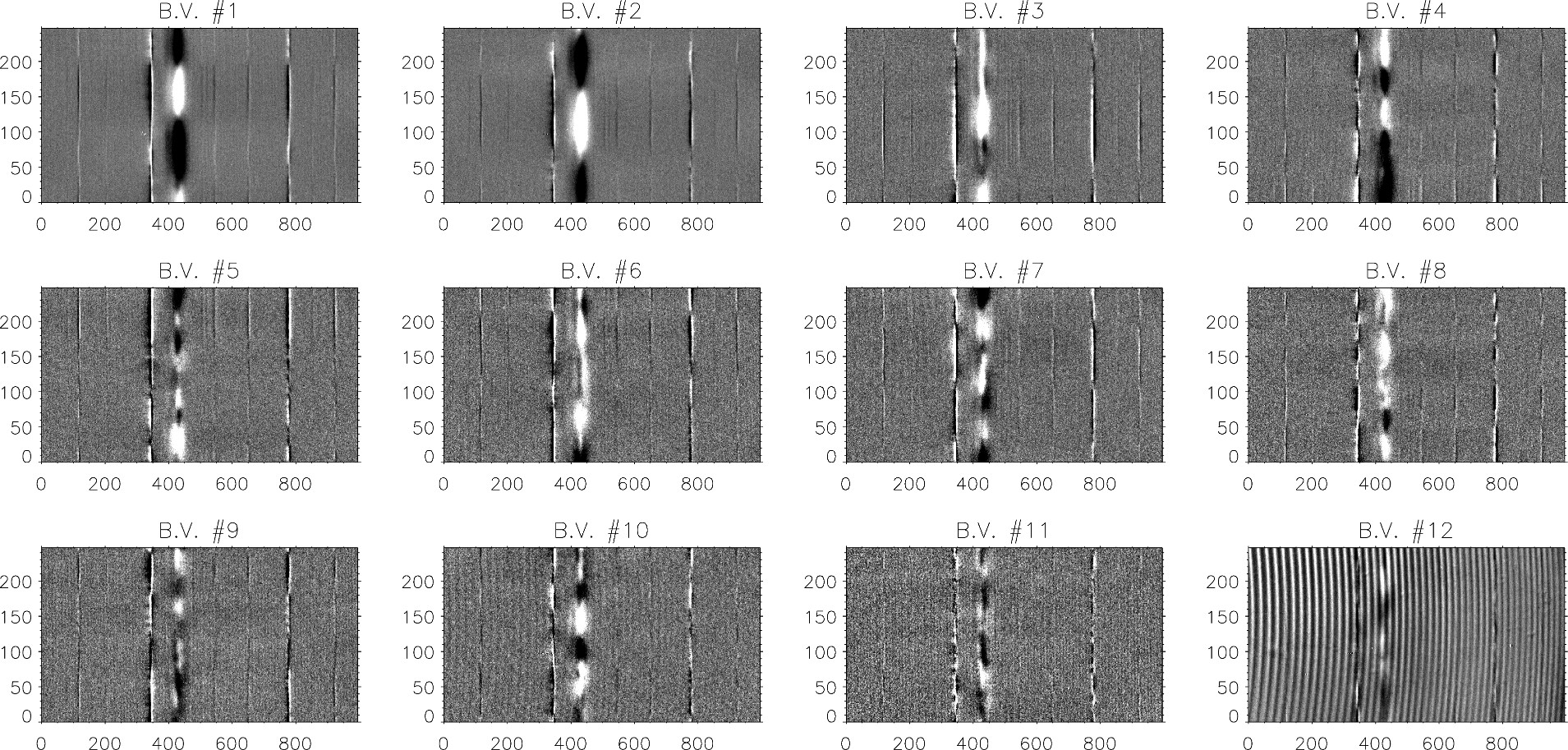}
\vbox{\medskip Transformed basis\medskip}
\caption{\label{fig:bases}
\emph{Top:} the TP-PCA basis $\{\bm{e}_j(x,y)\}_{j=1,\ldots,12}$ of a Stokes $V$ map with $N=12$ frames. Note the presence of strong fringes in the dominant basis vectors (B.V.; esp.\ \#1 and \#3). The same fringes affect to a minor degree multiple others elements of the basis. \emph{Bottom:} the transformed basis $\{\bm{\epsilon}_k(x,y)\}_{k=1,\ldots,12}$. Note that the fringes are very well confined to the subspace spanned by the last transformed basis vector. We also note that the residual target signal $\mathscr{S}$ in $\bm{\epsilon}_{12}(x,y)$ is visibly smaller than the one in both $\bm{e}_1(x,y)$ and $\bm{e}_3(x,y)$ of the original basis. Hence, reconstruction of the data set by dropping $\bm{\epsilon}_{12}(x,y)$ in the PCA reconstruction suppresses the fringes with less science data loss than by using the original basis without $\bm{e}_1(x,y)$ and $\bm{e}_3(x,y)$ (see Figure~\ref{fig:reconstr}).}
\end{figure}

\begin{figure}[t!]
\centering
\includegraphics[]{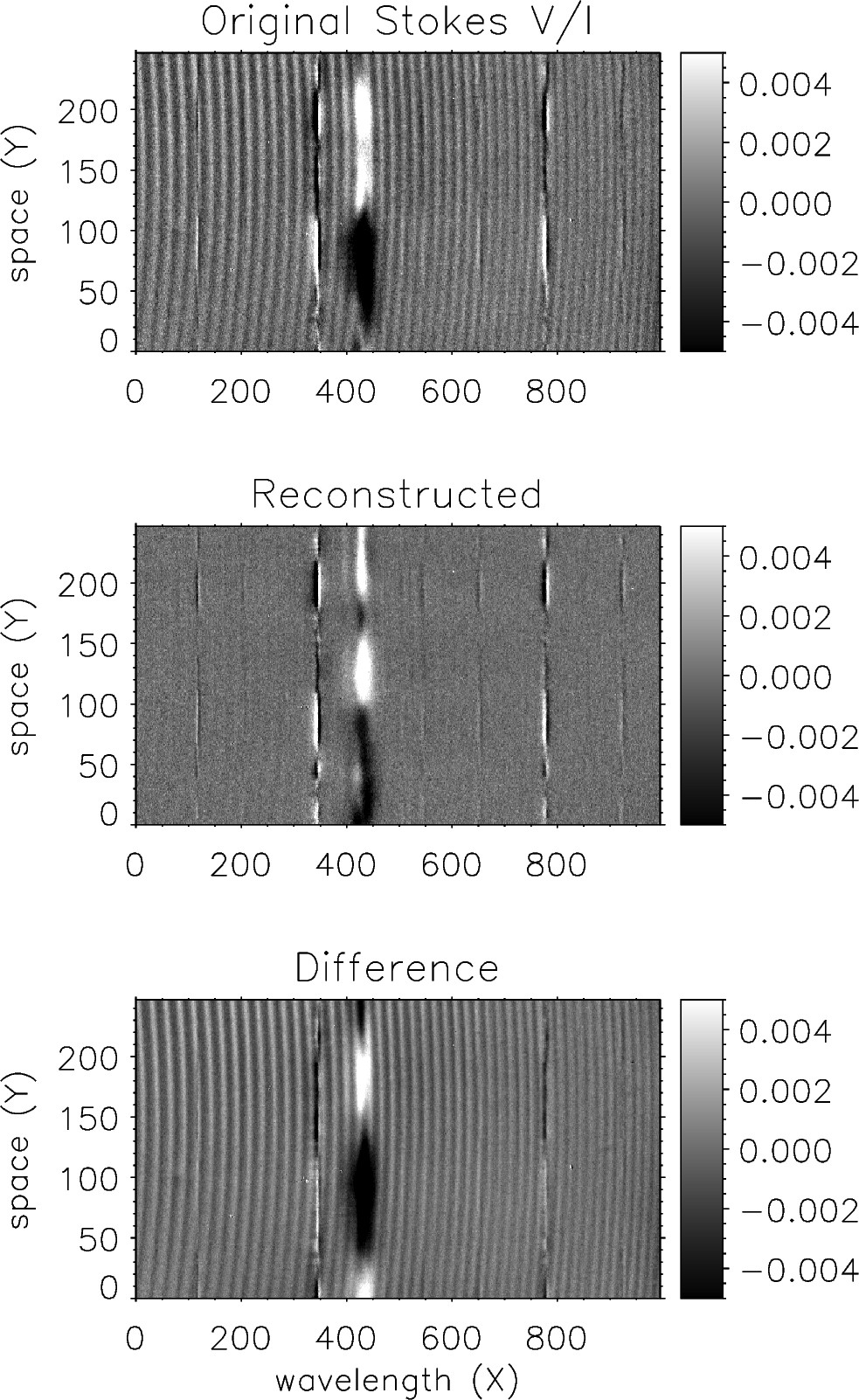}\kern 12pt
\includegraphics[]{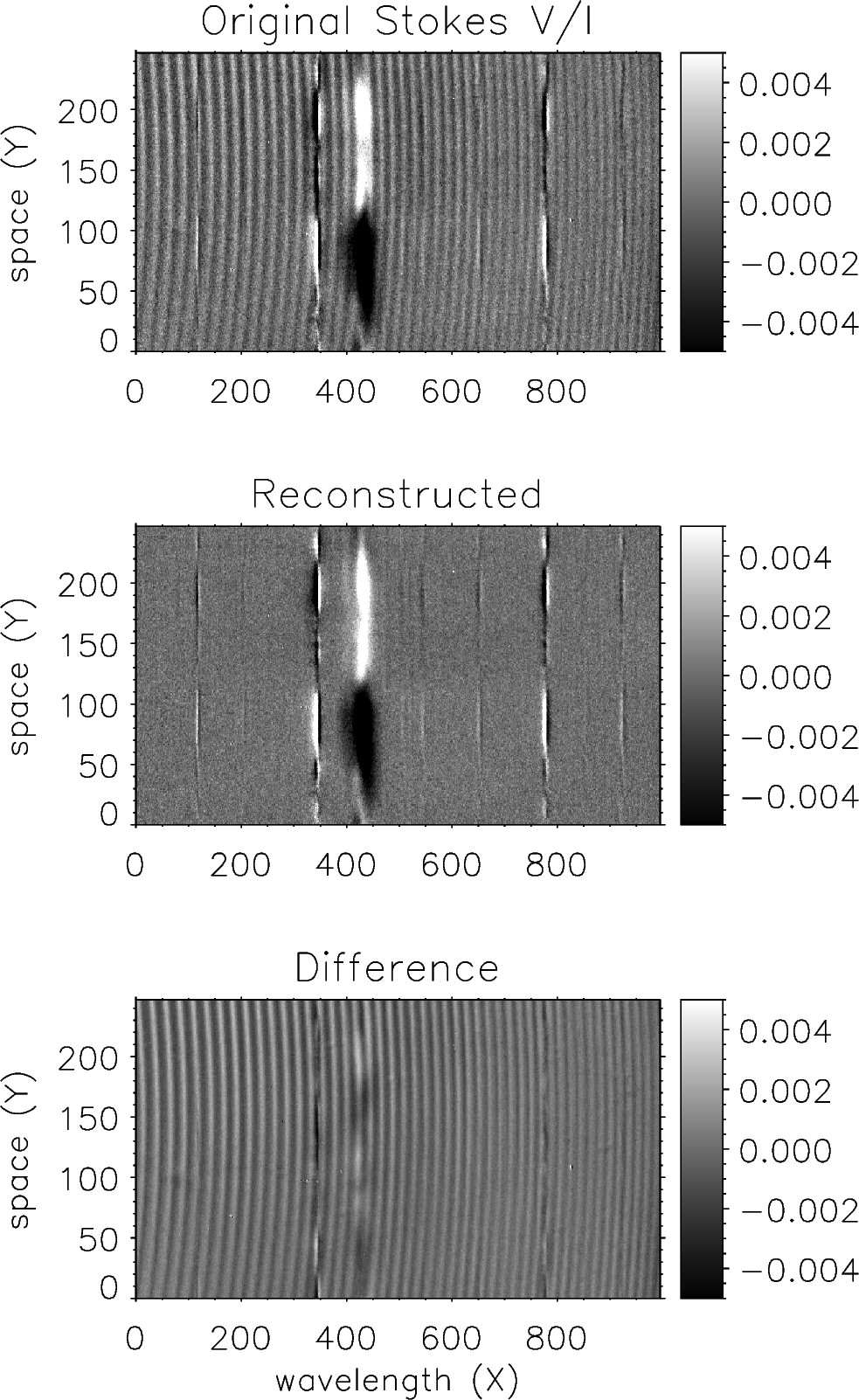}
\caption{\label{fig:reconstr}
Original (top), PCA-reconstructed (center), and residual of the reconstruction (bottom) for one frame of the same data set used for the creation of the bases of Figure~\ref{fig:bases}. \emph{Left:} using the original TP-PCA basis without $\bm{e}_1(x,y)$ and $\bm{e}_3(x,y)$. \emph{Right:} using the transformed basis without $\bm{\epsilon}_{12}(x,y)$.} 
\end{figure}

Because of the different physical origin of the spectral and instrumental signals (assuming that the instrument behaves as a linear system), the signals are in principle statistically independent, i.e., \emph{uncorrelated}. However, the type and conditions of the observations represented by the (limited) data set $\mathscr{D}$ might be such that this condition of uncorrelation is not fully realized in the data. For the development below, we assume that the set $\mathscr{D}$ indeed satisfies such condition,
which is expressed algebraically as
\begin{equation} \label{eq:uncorr}
\textrm{Corr}(\mathscr{S},\mathscr{F})=0\;.
\end{equation}
The $(i,j)$ element of this correlation matrix is the inner products
$\langle \mathscr{S}_i,\mathscr{F}_j \rangle$ (see App.~\ref{app:A}),
and so the condition (\ref{eq:uncorr}) of uncorrelation between $\mathscr{S}$ and $\mathscr{F}$ is equivalent to 
the orthogonality condition between the corresponding spaces, i.e., $\mathscr{F}=\mathscr{S}_\perp$ in $\mathscr{D}$.
This implies that $\mathscr{D}$ is the direct sum of $\mathscr{S}$ and $\mathscr{F}$, i.e., $\mathscr{D}=\mathscr{S}\oplus\mathscr{F}=\mathscr{S}\oplus\mathscr{S}_\perp$, and so the union of any two bases of $\mathscr{S}$ and $\mathscr{F}$ is also a basis of $\mathscr{D}$  \cite[see, e.g.,][Ch.~VII]{BM53}.
Therefore, there must exist an orthogonal transformation (isometry) $\bm{\mathsf{R}}$ of the PCA basis $\{\bm{e}_j(x,y)\}_{j=1,\ldots,N}$ into the basis
%
$\{\bm{\epsilon}_k(x,y)\}_{k=1,\ldots,N}$,
\begin{equation} \label{eq:newbasis}
\bm{\epsilon}_k(x,y)=\sum_{j=1}^N \mathsf{R}_{kj}\,\bm{e}_j(x,y)\;,
\end{equation}
such that $\mathscr{S}$ is generated (after a possible re-ordering of the basis) by the subset 
$\{\bm{\epsilon}_k(x,y)\}_{k=1,\ldots,n}$, for some $n<N$, whereas $\mathscr{F}=\mathscr{S}_\perp$ is generated by the complementary subset $\{\bm{\epsilon}_k(x,y)\}_{k=n+1,\ldots,N}$.

This allows us to rewrite eq.~(\ref{eq:expans.corr}) as 
\begin{equation} \label{eq:expans.d1}
\mathscr{D}_i(x,y)=\sum_{k=1}^N \gamma_{ik}\,\bm{\epsilon}_k(x,y)\;,
\end{equation}
where
\begin{equation}
\gamma_{ik}=\sum_{j=1}^N c_{ij}\,\mathsf{R}_{kj}=\sum_{j=1}^N c_{ij}\,\mathsf{R}^T_{jk}\;.
\end{equation}
Similarly, for the de-fringed data, we have
\begin{equation} \label{eq:expans.s}
\mathscr{S}_i(x,y)=\sum_{k=1}^n \gamma_{ik}\,\bm{\epsilon}_k(x,y)\;.
\end{equation}

The objective of the de-fringing process by 2D PCA is therefore to determine the isometry $\bm{\mathsf{R}}$ that produces a basis set 
$\{\bm{\epsilon}_k(x,y)\}_{k=1,\ldots,N}$ with the properties given above. The realizability of this objective may in practice be hindered by the observing conditions, the nature of the target, and the limitedness of the data set, as these can prevent the full realization of the uncorrelation condition (\ref{eq:uncorr}) within a specific data set $\mathscr{D}$.

Despite this practical limitation, it is always possible to devise an algorithm that determines an optimal set of rotations between two any eigenfeatures with the purpose of eliminating the unwanted signal $\mathscr{F}$ from one of them. In App.~\ref{app:B}, we list the steps of a simple procedure that can easily be automated. By recursively applying that algorithm through the basis
$\{\bm{e}_j(x,y)\}_{j=1,\ldots,N}$ of $\mathscr{D}$, the unwanted signal can thus be confined to a \emph{minimal} (i.e., no longer reducible) set of transformed basis vectors 
$\{\bm{\epsilon}_k(x,y)\}_{k=n+1,\ldots,N}$. Any residual spectral signal in such set practically quantifies the departure of the uncorrelation condition (\ref{eq:uncorr}) from being fully realized within the data set $\mathscr{D}$.

It is important to remark that, in general, the transformed basis 
$\{\bm{\epsilon}_k(x,y)\}_{k=1,\ldots,N}$ no longer represents a basis of PCs of the data set, as it no longer corresponds to the original matrix $\bm{\mathsf{U}}$ of eigenvectors of the correlation matrix $\mathrm{Corr}(\mathscr{D},\mathscr{D})$ via eq.~(\ref{eq:2D_basis.corr}). In particular, this implies that the original set of eigenvalues cannot be used to quantify the relative contributions of the transformed basis vectors to the data (see end of Sect.~\ref{sec:YPCA}).


\section{Examples} \label{sec:examples}

The concepts presented in the previous section are well demonstrated by Figures~\ref{fig:bases} and \ref{fig:reconstr}. For these, we considered a subset of 12 frames taken from a spectro-polarimetric map of a solar active region in the \ion{He}{1} 1083\,nm multiplet, observed in 2011 with the Facility Infra-Red Spectro-polarimeter \cite[FIRS;][]{Ja10} at the Dunn Solar Telescope (DST) of the National Solar Observatory on Sacramento Peak (NSO/SP; Sunspot, NM). The figures show results of the application of the TP-PCA approach to the Stokes $V$ (circular polarization) signal.


\begin{figure}[t!]
\centering
\includegraphics[]{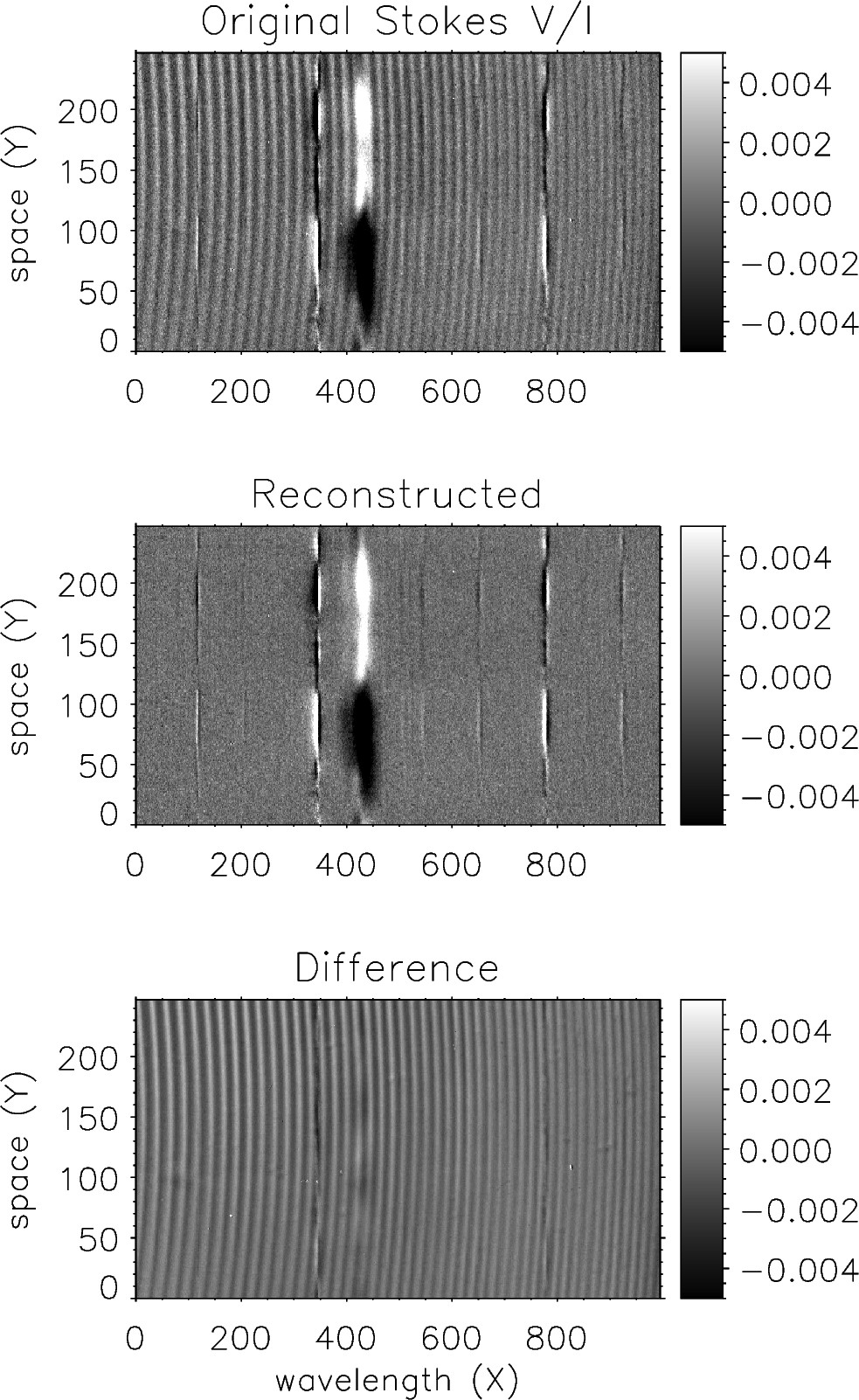}\kern 12pt
\includegraphics[]{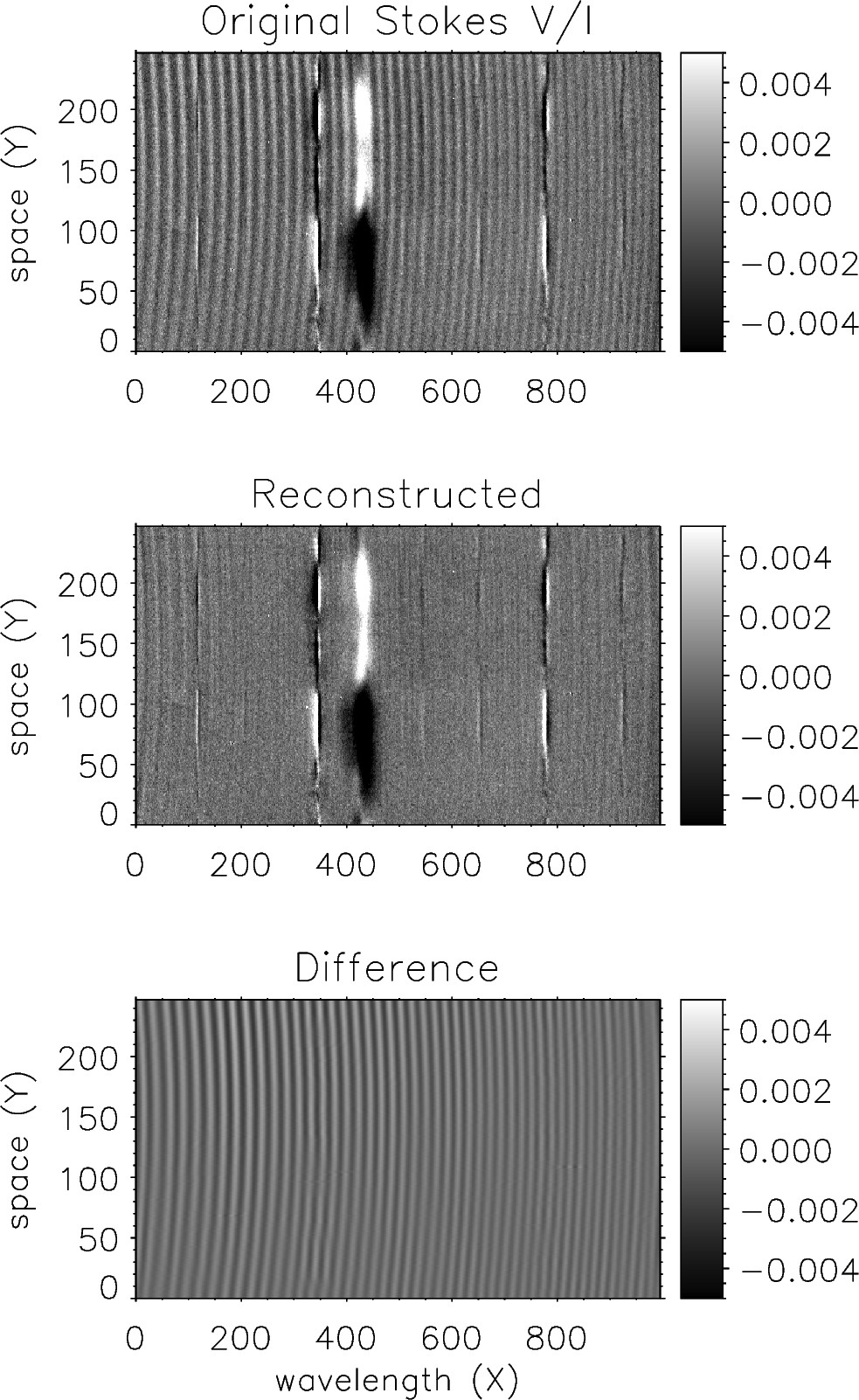}
\caption{\label{fig:reconstr.32-FF}
Reconstruction of the same map step as in Figure~\ref{fig:reconstr}, but using a basis with 32 vectors. \emph{Left:} dropping $\bm{\epsilon}_{32}(x,y)$ from the reconstruction (similarly to the right side of Figure~\ref{fig:reconstr}). \emph{Right:} including $\bm{\epsilon}_{32}(x,y)$ in the reconstruction, but after running a Fourier filtering of that basis vector, in order to remove the confined fringes after the basis transformation.}
\end{figure}

Figure~\ref{fig:bases} shows the original TP-PCA basis $\{\bm{e}_j(x,y)\}_{j=1,\ldots,N}$ (top) and the transformed basis
$\{\bm{\epsilon}_k(x,y)\}_{k=1,\ldots,N}$ (bottom), the latter being derived by the recursive application of the algorithm described in App.~\ref{app:B}.  
The original basis shows the presence of fringes in several basis vectors, most notably in the eigenfeatures $\bm{e}_1(x,y)$ and $\bm{e}_3(x,y)$. The determination of the PCA basis relies on the Singular Value Decomposition (SVD) of the correlation matrix, which produces an ordered set of eigenfeatures according to the decreasing importance of their contribution to the data set. Therefore, the presence of strong spectral signals in low-order eigenfeatures is of great concern, since a simple elimination of those eigenfeatures in order to suppress the fringe signal would cause a significant loss of the spectral signal targeted by the science. By applying the optimal rotation algorithm described in App.~\ref{app:B}, the fringe signal in each eigenfeature is moved down the transformed basis $\{\bm{\epsilon}_k(x,y)\}_{k=1,\ldots,N}$, until it is completely constrained within the last few basis vectors.
Figure~\ref{fig:bases} demonstrates that, for this specific data set, the rotation algorithm is particularly effective, practically limiting the presence of the unwanted signal $\mathscr{F}$ to just the last basis vector $\bm{\epsilon}_{12}(x,y)$. 

While the basis transformation generally modifies the ``importance'' ordering of the original PCA basis, the fundamental result of the transformation is to accomplish the reduction---and ideally, the full suppression---of the targeted spectral signal in the fringe eigenfeatures, essentially realizing the process of orthogonalization of $\mathscr{S}$ and $\mathscr{F}$ within the basis of the data set $\mathscr{D}$. Of course, perfect orthogonality (and hence, uncorrelation) in the data set is implied only when the transformed basis vectors generating $\mathscr{F}$ are completely devoid of the spectral signal $\mathscr{S}$. This is certainly not the case for the transformed basis of Figure~\ref{fig:bases}. However, the residual spectral signal in the last basis vector $\bm{\epsilon}_{12}(x,y)$ has been significantly reduced.

\begin{figure}[t!]
\centering
\includegraphics[width=7in]{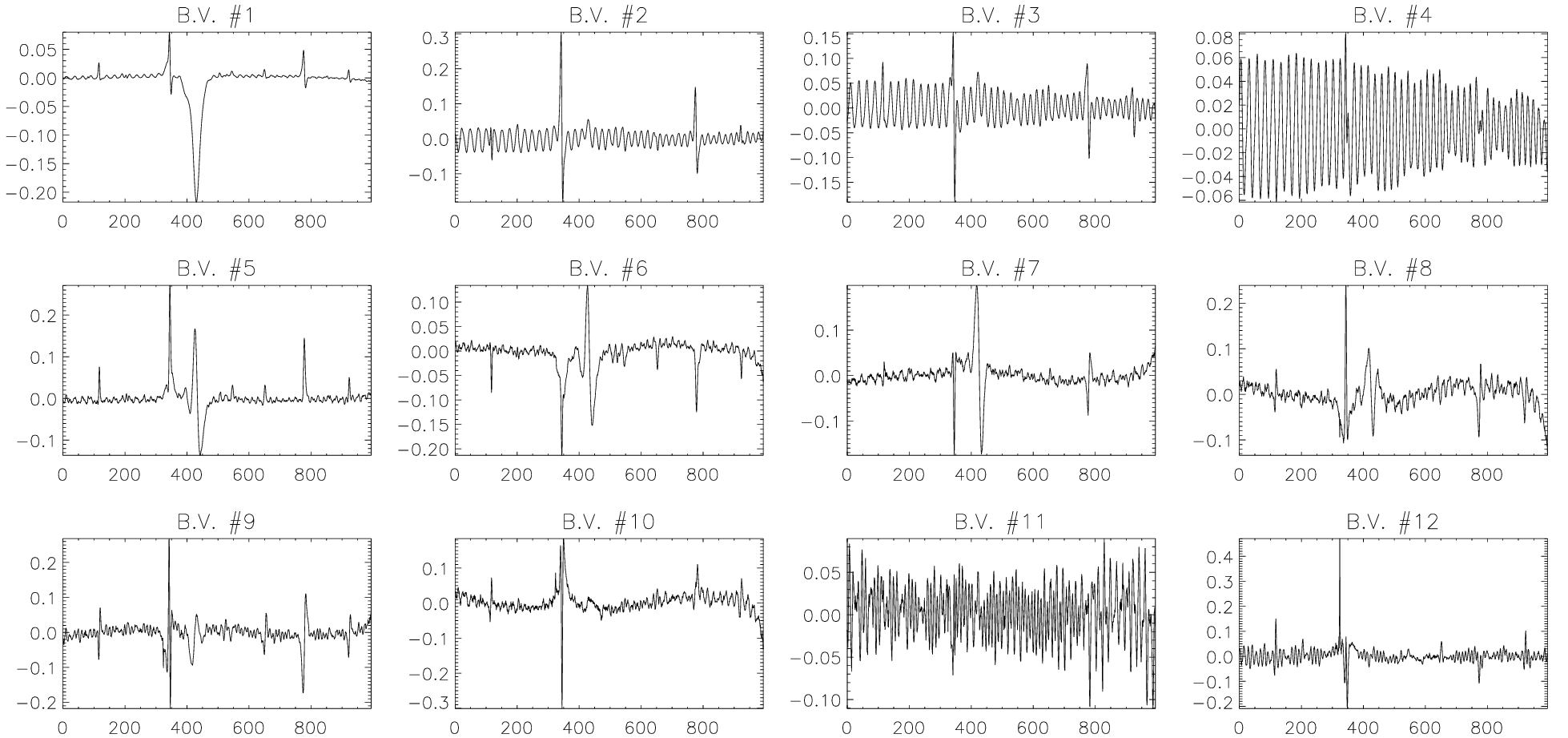}
\vbox{\medskip Original PCA basis\bigskip}
\includegraphics[width=7in]{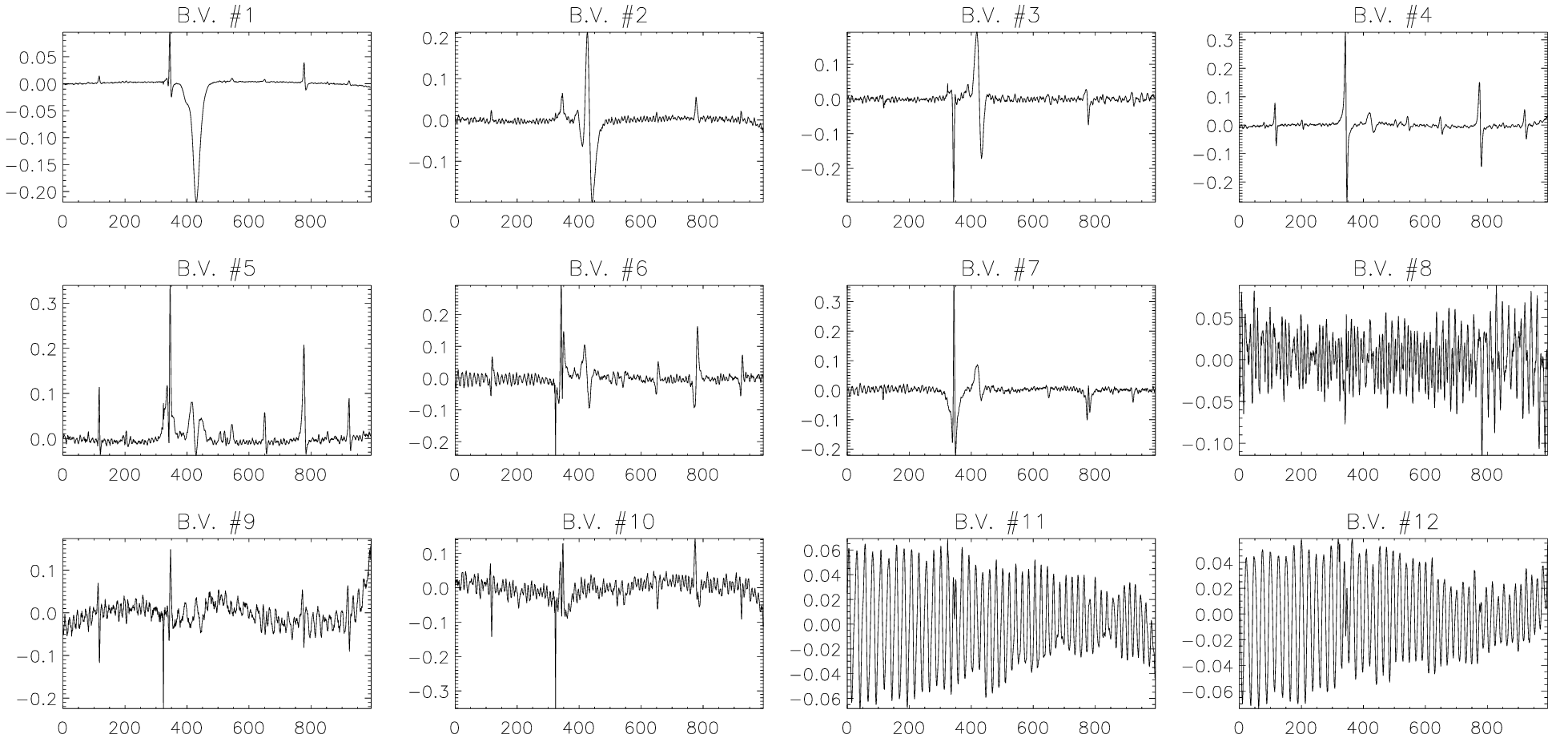}
\vbox{\medskip Transformed basis\bigskip}
\caption{\label{fig:bases.1D}
\emph{Top:} the Y-PCA basis $\{\bm{e}_j(x)\}_{j=1,\ldots,12}$ for the same Stokes $V$ map with $N=32$ frames used for Figure~\ref{fig:reconstr.32-FF}. Note the evident presence of fringes in the eigenprofiles 2 to 4. The same fringes affect to a minor degree multiple others elements of the basis, including $\bm{e}_1(x)$. \emph{Bottom:} the transformed basis 
$\{\bm{\epsilon}_k(x)\}_{k=1,\ldots,12}$. Note that the fringes have been very well confined to the subspace spanned by the last two basis vectors, and significantly suppressed elsewhere.}
\end{figure}

The above examples were restricted to a small set of Stokes map's frames for practical reasons, and it can be expected that the reduced statistical significance of such a set importantly impacts the realization of the uncorrelation condition (\ref{eq:uncorr}). In typical science applications, instead, the number of frames is much larger (of order $10^2$), and this should favor the realizability of the condition (\ref{eq:uncorr}), \emph{if the Stokes signals of the observed target are sufficiently varied across the map.}
This is well demonstrated by Figure~\ref{fig:reconstr.32-FF} (left), which shows how the quality of fringe removal in the previous example improves when the PCA basis is determined using 32 frames in the map, instead of the 12 used for Figure~\ref{fig:reconstr}.

\begin{figure}[t!]
\centering
\includegraphics[]{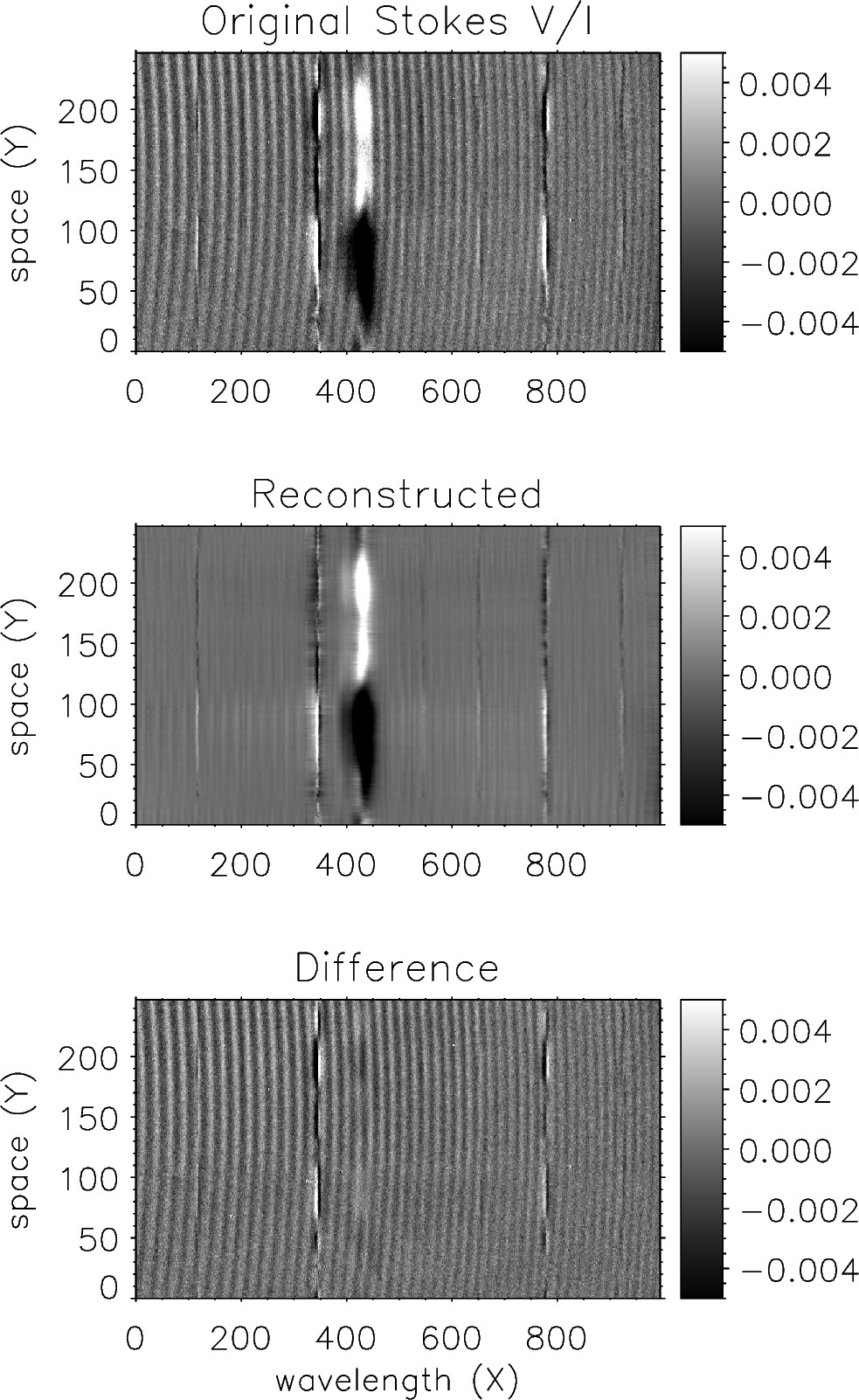}\kern 12pt
\includegraphics[]{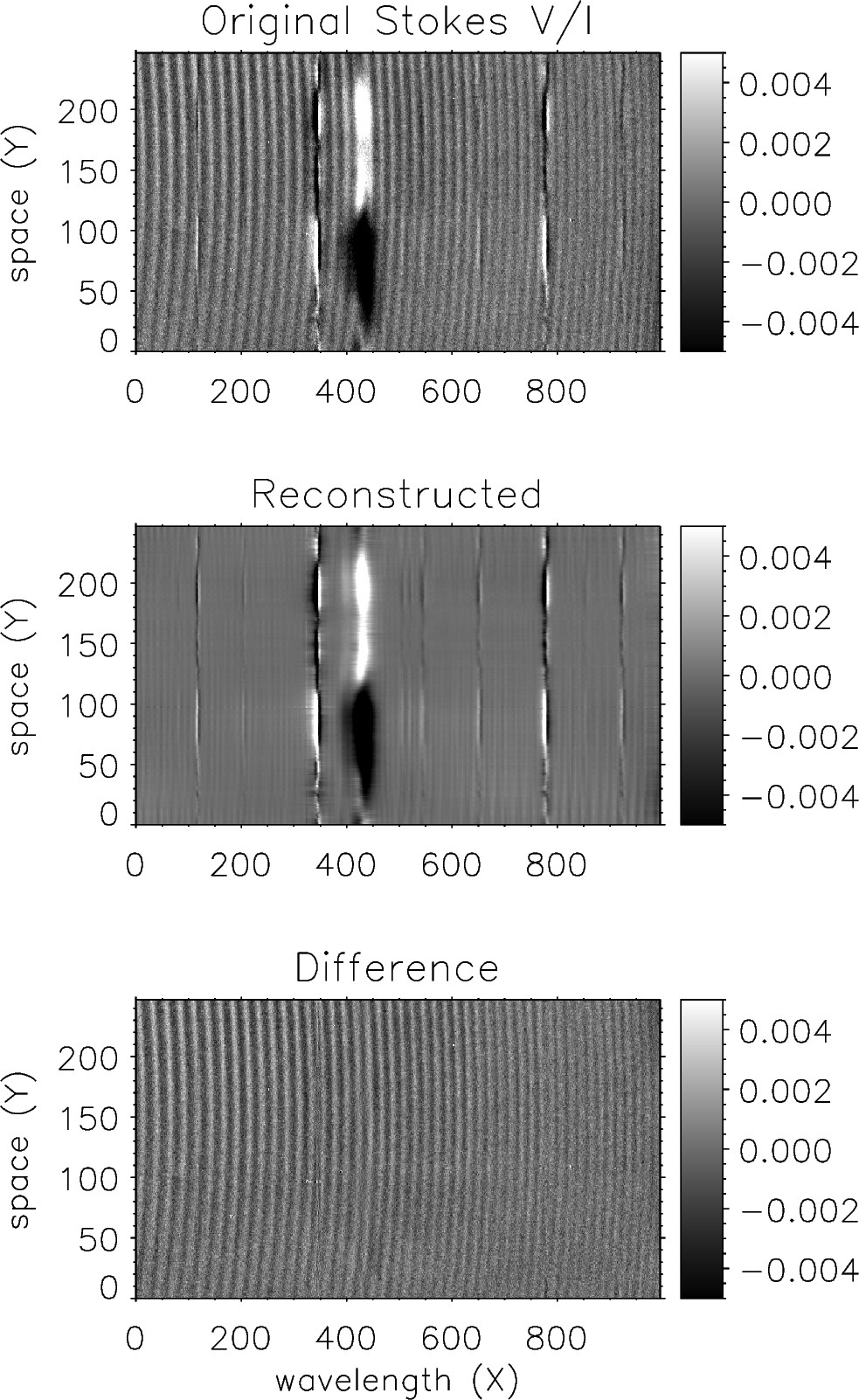}
\caption{\label{fig:reconstr.1D}
Original (top), Y-PCA reconstructed (center), and residual of the reconstruction (bottom) for a specific frame of the data set used for the creation of the bases of Figure~\ref{fig:bases.1D}. \emph{Left:} using the original Y-PCA basis without $\bm{e}_2(x)$, $\bm{e}_3(x)$, and $\bm{e}_4(x)$. \emph{Right:} using the rotated basis without $\bm{\epsilon}_{11}(x)$ and $\bm{\epsilon}_{12}(x)$. We note the significantly reduced loss of spectral signal with the rotated basis (especially for the two \ion{Si}{1} lines), despite a comparable suppression of the dominant fringes.}
\end{figure}

When the residual spectral signal in the set of transformed basis vectors containing the fringes is important, one can attempt to recover such signal rather than drop it from the reconstruction together with the unwanted fringes. A direct way for doing this is by Fourier filtering of the fringe component(s) from the relevant basis vectors. Once the fringes are removed from a basis vector, this can be added back to the basis set used for the data reconstruction, rather than being eliminated as done in the previous examples.
Figure~\ref{fig:reconstr.32-FF} (right) shows an example of such manipulation of basis elements by Fourier filtering. In the rotated basis of 32 elements, the last two showed visible fringes, particularly $\bm{\epsilon}_{32}(x,y)$, with non negligible contribution from the targeted spectral signal. Applying Fourier filtering to the last two basis vectors, that signal can be added back to the reconstruction, producing a residual without practically any science data loss (see right panels of Figure~\ref{fig:reconstr.32-FF}).

It is important to observe that the suppression of any signal from a basis vector, e.g., by Fourier filtering, breaks the orthonormality of the basis, which therefore can no longer be used as a projection vector set for the data. However, it is possible to reconstruct the data by using the same coefficient matrix $\bm{\mathsf{C}}$ for the transformed basis 
$\{\bm{\epsilon}_i\}_{i=1,\ldots,N}$ prior filtering. This is mathematically equivalent to performing the data reconstruction by excluding the basis vectors affected by the fringes, and then apply the Fourier filtering to the difference image, in order to retrieve the residual target signal from it. The downside of the second approach is that the residual must in principle be Fourier filtered independently for each map step (although it is easy to devise an automated way to do so for an entire map), whereas by directly Fourier filtering the basis set, the residual correction is automatically applied to the entire Stokes map.

In Sect.~\ref{sec:FF}, we compare this \emph{selective} Fourier filtering of fringe PCs or difference images with the more traditional application of Fourier filtering directly to a frame of the Stokes map.


\subsection{Application to the Y-PCA method}
\label{sec:YPCA}

The technique of orthogonal transformations of the PCA basis, in order to confine the polarization fringes components to an orthogonal subspace of the spectral signal, can also be employed to improve the performance of the Y-PCA method considered by \cite{Ca12}.

In order to test its performance, we employed the same Stokes $V$ map with 32 scan steps used for the results of Figure~\ref{fig:reconstr.32-FF}. The Y-PCA algorithm implies the creation of the data correlation matrix \cite[cf.\ eq.~(1) of][]{Ca12}
\begin{equation} \label{eq:corr.1D}
\textrm{Corr}(\mathscr{D},\mathscr{D})=\frac{1}{N}\sum_{i=1}^N 
\mathscr{D}_i(x,y)^T \mathscr{D}_i(x,y)\;.
\end{equation}
The above definition implies contraction over the spatial dimension $Y$, and leads to a $N_X\times N_X$ correlation matrix, with $N_X$ eigenprofiles in the resulting PCA basis.

\begin{figure}[t!]
\centering
\includegraphics[width=7in]{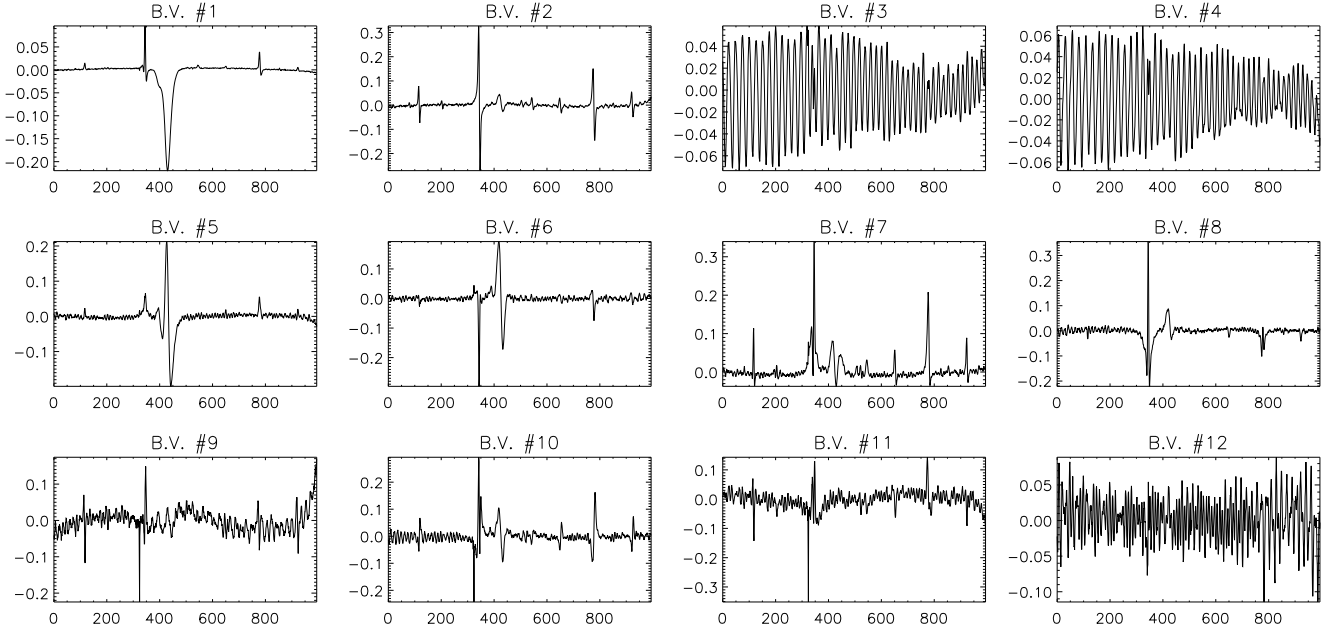}
\caption{\label{fig:reord}
The transformed basis of Figure~\ref{fig:bases.1D} after reordering of the basis vectors according to the weight of their relative contributions to the map data.}
\end{figure}

In the case of the Stokes maps considered in the previous examples, $N_X=995$, but in Figure~\ref{fig:bases.1D} we only consider the first 12 basis vectors. The fringes are easily recognizable, and in order to suppress them by using the decomposition of the Stokes map on the original PCA basis (top half of Figure~\ref{fig:bases.1D}), one must drop at least $\bm{e}_2(x)$, $\bm{e}_3(x)$, and $\bm{e}_4(x)$ from the data reconstruction. The result of this operation is shown on the left of Figure~\ref{fig:reconstr.1D}. Qualitatively, it fares comparably to the TP-PCA reconstruction using the full basis of 32 vectors after rotation (left side of Figure~\ref{fig:reconstr.32-FF})---slightly better for \ion{He}{1} 1083\,nm, but visibly worse for the two \ion{Si}{1} lines. However, using the transformed basis (bottom half of Figure~\ref{fig:bases.1D}), where the fringes are very well confined to the last two basis vectors $\bm{\epsilon}_{11}(x)$ and $\bm{\epsilon}_{12}(x)$, the suppression of the fringes in the reconstructed data by dropping those two basis vectors improves significantly, becoming qualitatively comparable to the TP-PCA reconstruction after Fourier filtering of the fringe-carrying basis vectors (cf.\ the right sides of Figs.~\ref{fig:reconstr.32-FF} and \ref{fig:reconstr.1D}). Incidentally, we note the complete absence of random ``noise'' in the reconstructed images of Figure~\ref{fig:reconstr.1D}, because of the low-pass cut implied by considering only the lowest-order 12 basis vectors out of the full set of $N_X=995$.

\begin{figure}
\centering
\includegraphics[width=.49\hsize]{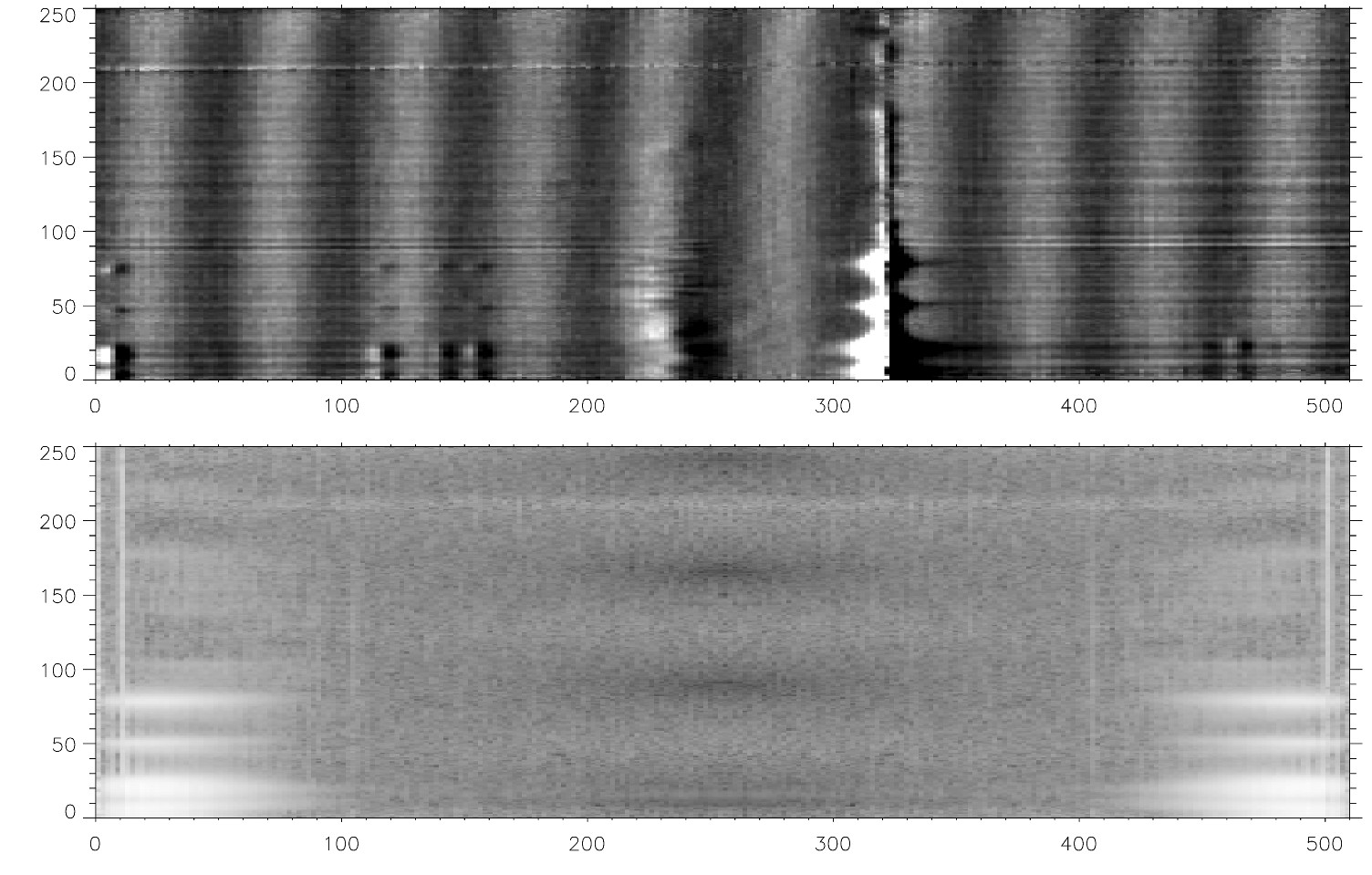}\kern 6pt
\includegraphics[width=.49\hsize]{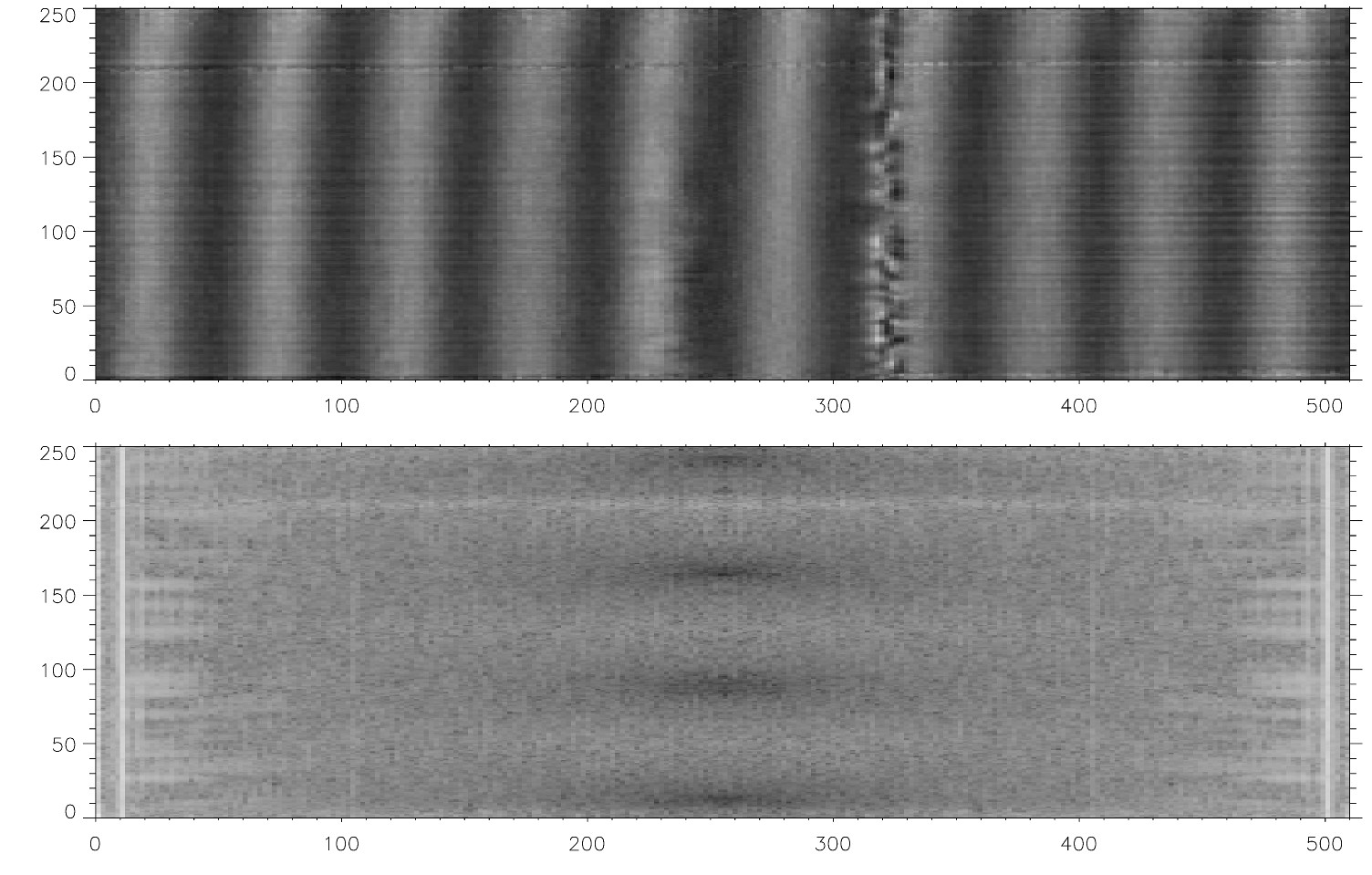}
\caption{\label{fig:AC}
\emph{Left:} A Stokes map frame (top), and its corresponding spectral autocorrelation image (bottom). The frequency range of the dominant fringes is clearly visible as two vertical strips in Fourier space, at around $X=10$ and $X=400$, stretching along the entire spatial dimension. Note the significant amplitude overlap of these fringes with the spectral signals in several spatial regions below $Y=100$. \emph{Right:} The residual of the same frame after PCA reconstruction by excluding the fringe affected basis vectors (top), and its corresponding spectral autocorrelation image (bottom). We note the much improved isolation of the fringe contribution in Fourier space, due to the overall suppression of the spectral signal, which allows for an efficient filtering of the fringes in the residual image.}
\end{figure}

\begin{figure}[t!]
\centering
\includegraphics[]{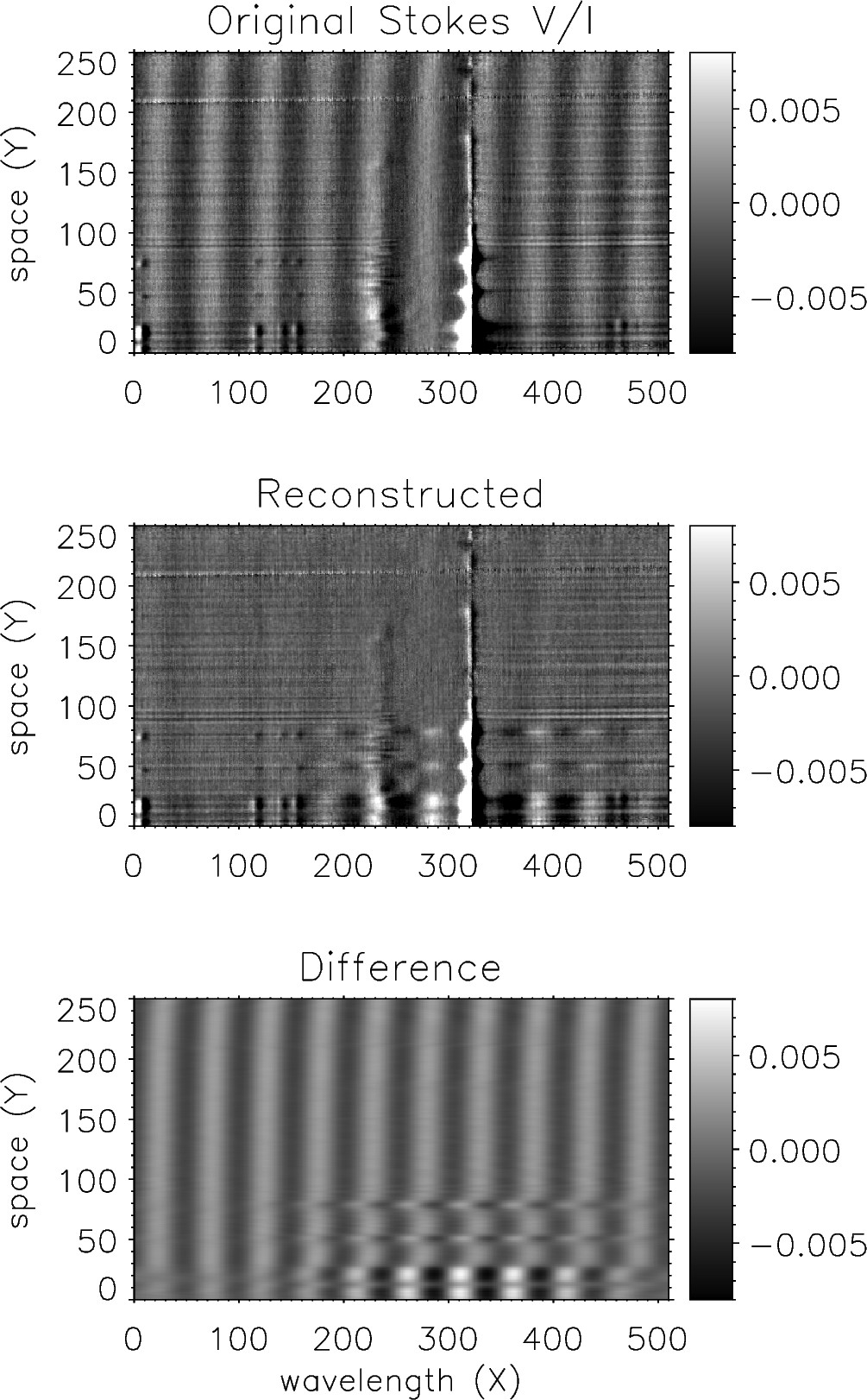}\kern 12pt
\includegraphics[]{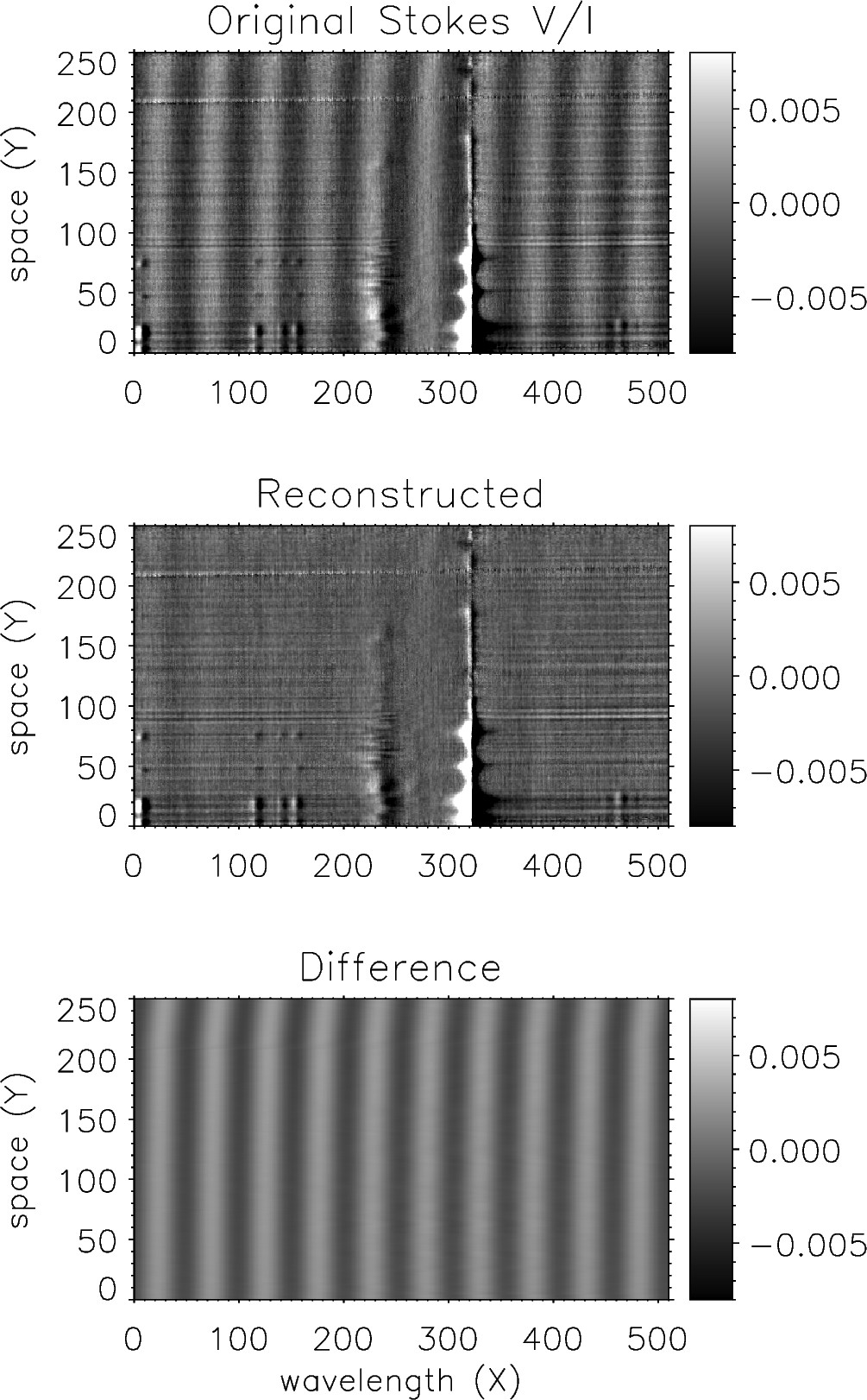}\vskip 6pt
\includegraphics[width=.495\hsize]{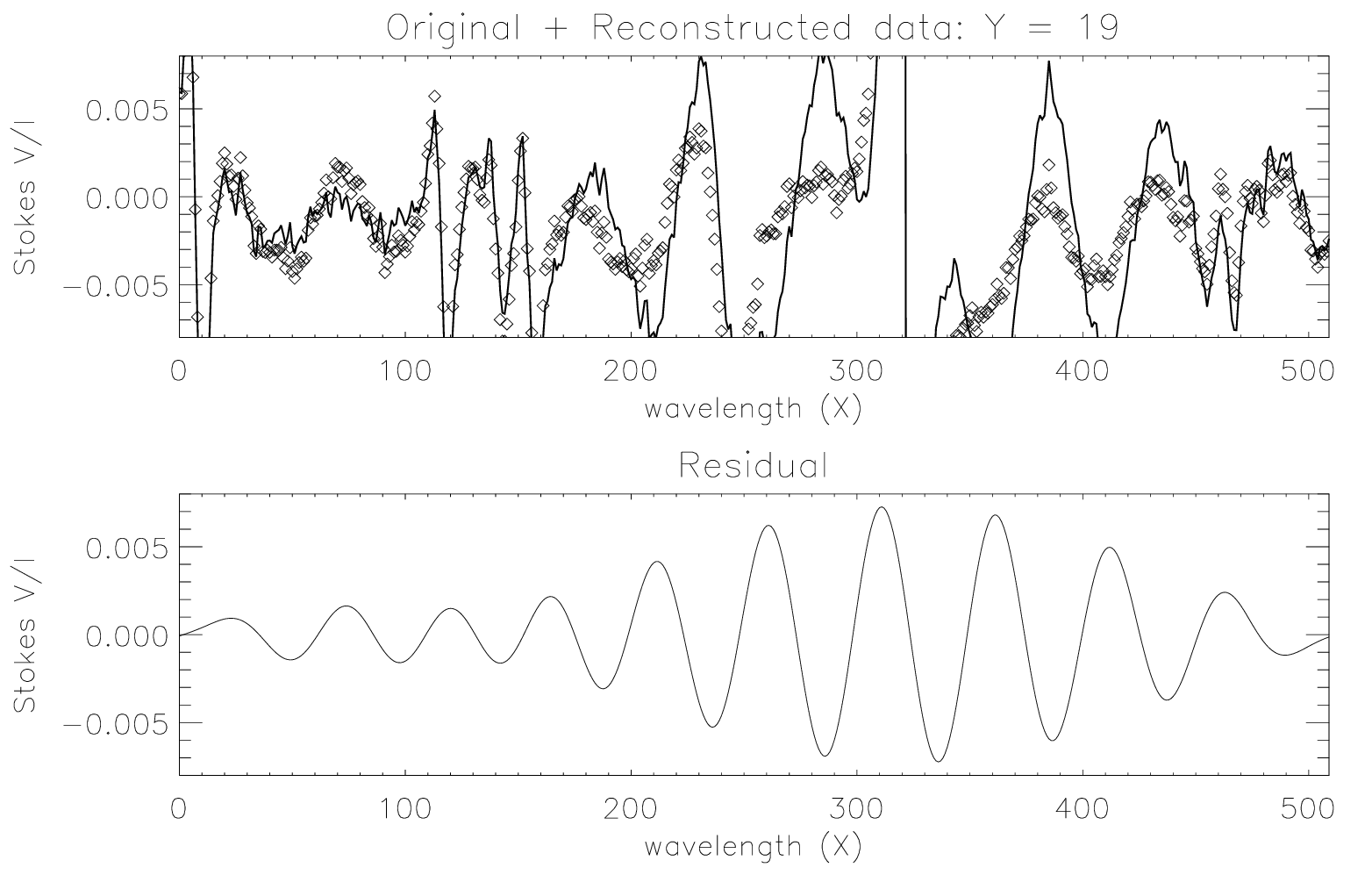}\kern 6pt
\includegraphics[width=.495\hsize]{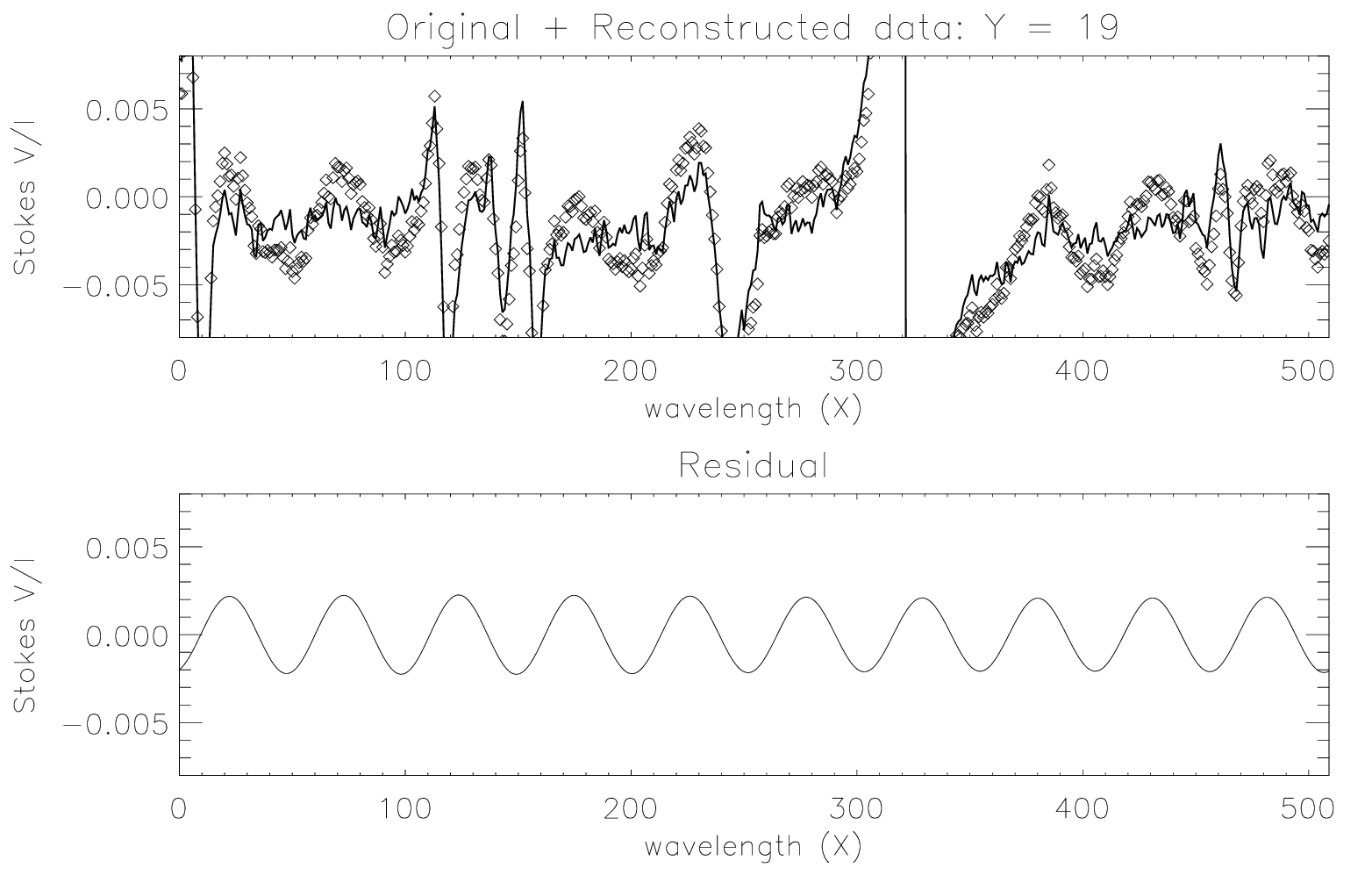}
\caption{\label{fig:reconstr.1083}
Original (1st row), reconstructed (2nd row), and residual of the reconstruction (3rd row) for one frame of a Stokes $V$ map around the \ion{He}{1}  1083\,nm line. The map (frequency ordered) was taken with the DST/SPINOR instrument, and contains a total of 100 frames. \emph{Left:} Stokes signal reconstructed after a direct Fourier filtering of the frame. \emph{Right:} Stokes signal reconstructed after a selective Fourier filtering of the contribution from the fringe basis vectors, after a TP-PCA of the full map, and successive transformation of the PCA basis. The bottom two rows show the profiles at $Y=19$ in the corresponding reconstructed frames. \emph{Top row:} original data (diamonds) and the reconstructed profile (continuous line). \emph{Bottom row:} filtered fringes. We note the significant spurious signal introduced by the direct Fourier filtering of the frame shown on the left.}
\end{figure}

\begin{figure}[t!]
\centering
\includegraphics[]{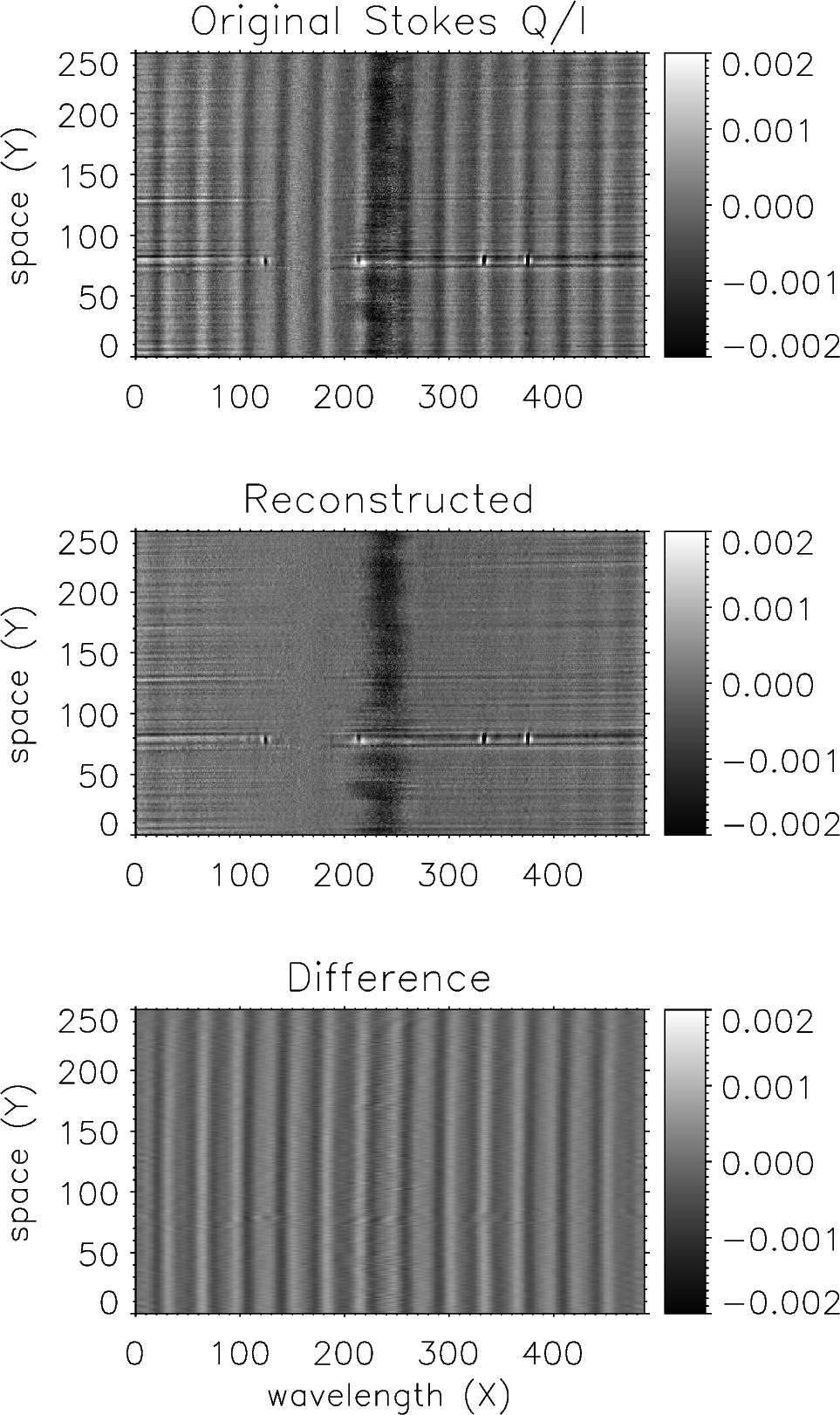}\kern 12pt
\includegraphics[]{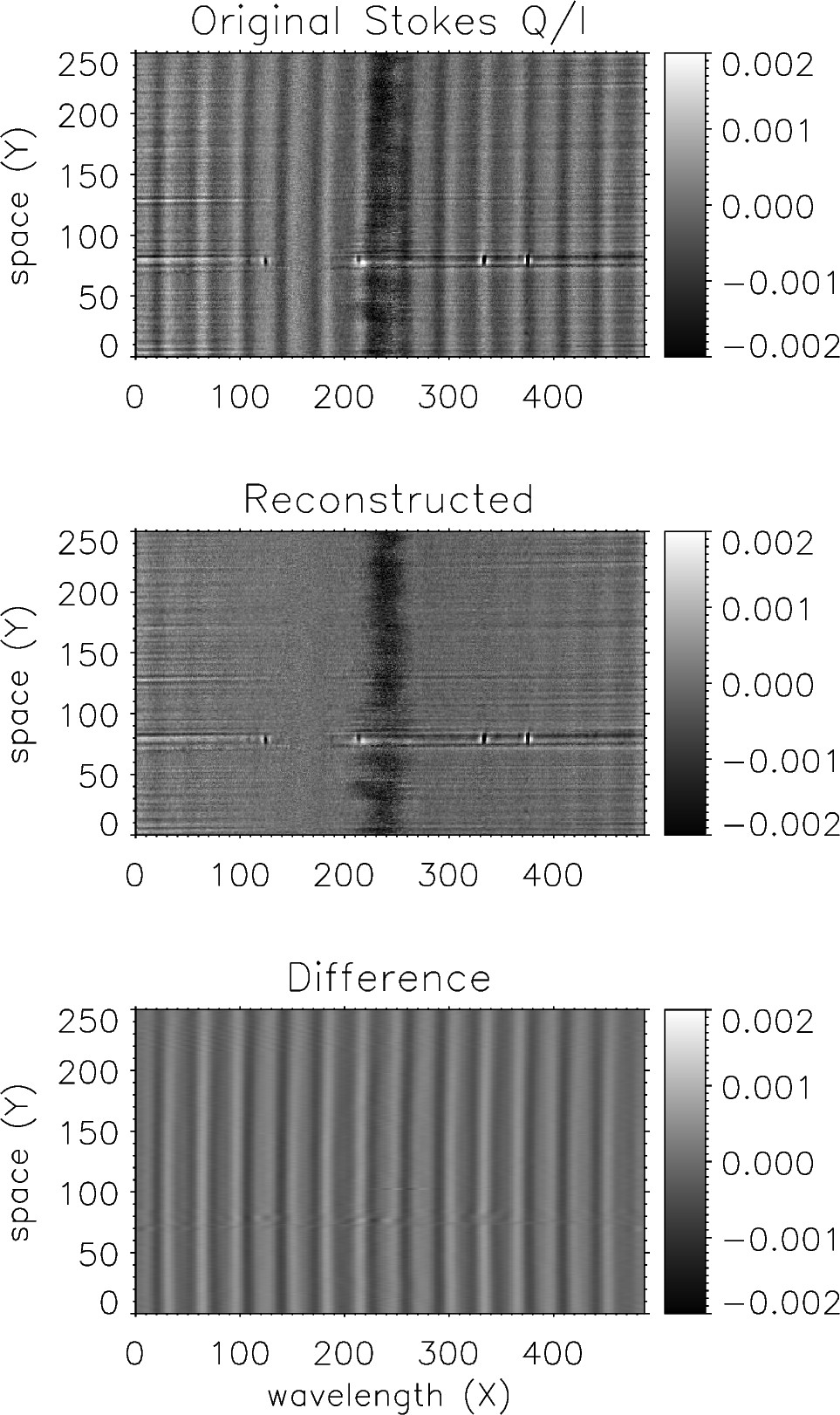}\vskip 12pt
\includegraphics[width=.495\hsize]{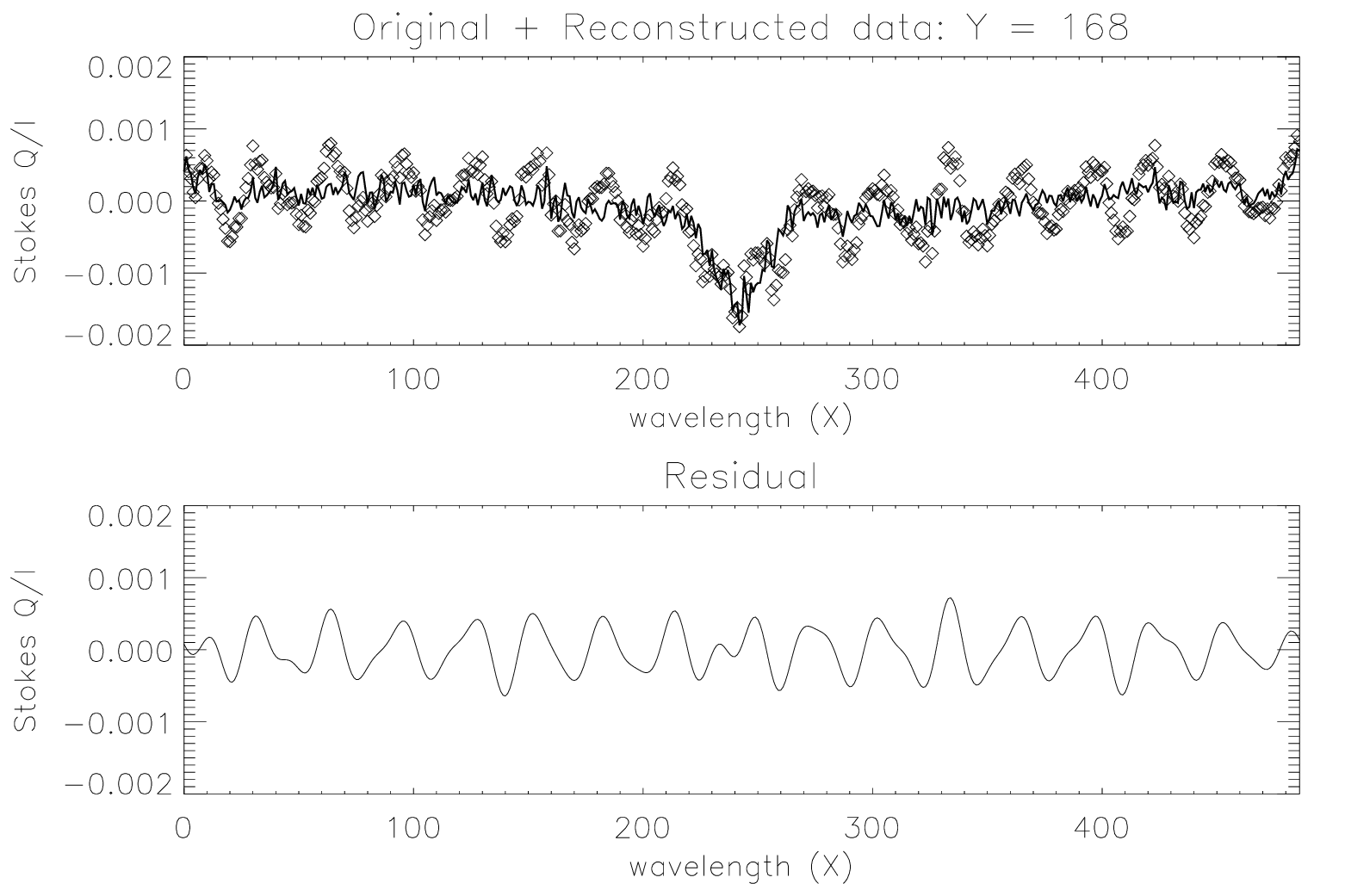}\kern 6pt
\includegraphics[width=.495\hsize]{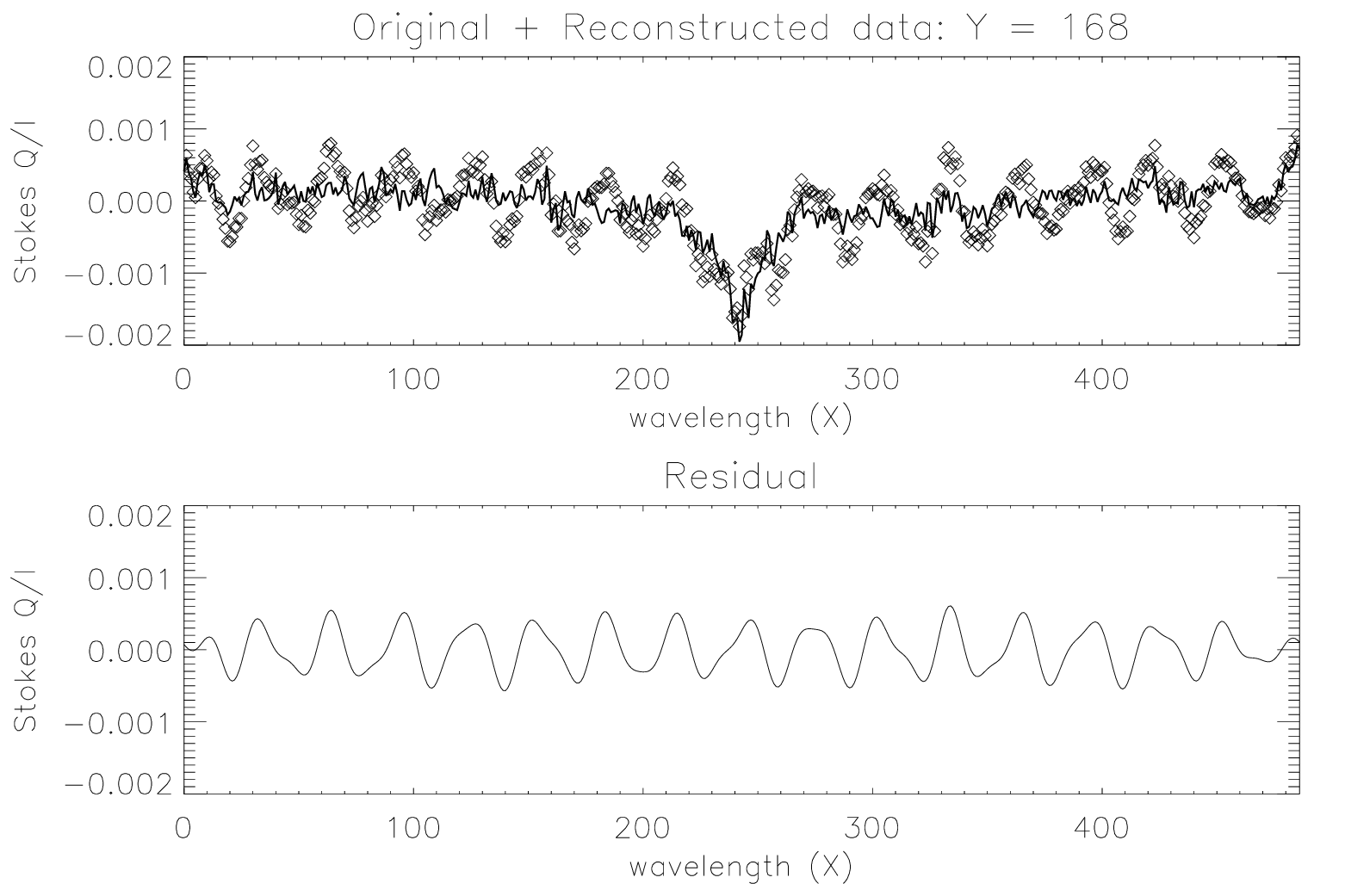}
\caption{\label{fig:reconstr.Ha}
Same as Figure~\ref{fig:reconstr.1083} for one frame of a Stokes $Q$ map of the \ion{H}{1} H$\alpha$ line at 656\,nm. The map (frequency ordered) was taken with the DST/SPINOR instrument, and contains a total of 60 frames. For this data set, direct Fourier filtering (left) produces a very good result, with only minor differences with respect to selective Fourier filtering of the contribution from the dominant fringe basis vectors (in this case $\bm{e}_1$, $\bm{e}_2$, and $\bm{e}_6$) shown at the right.}
\end{figure}

This test confirms the results of \cite{Ca12} about the applicability of the Y-PCA method to the Stokes de-fringing problem. However, depending on the particular data set, the state of polarization, and the instrument properties, having access to different methods for analyzing the data can help optimize the de-fringing process.

Typically, compared to Y-PCA, the TP-PCA method manages to confine the fringe subspace into a smaller number of basis vectors (i.e., $\mathscr{F}$ has a lower dimensionality), which are therefore easier to identify and manage for the data reconstruction, whether it is a matter of excluding them from the transformed basis, or of applying Fourier filtering to them in order to retrieve some residual spectral signal. One downside of both methods, when a basis rotation is applied, is the fact that the basis transformation scrambles the \emph{importance ordering} of the original PCA basis. For the sake of reducing data loss, one always tries to minimize the number of basis vectors of $\mathscr{F}$ that must be further handled, whether by simply dropping them from the data reconstruction or by Fourier filtering. Therefore, lack of information on the relative importance of the contributions of the transformed basis vectors to the data can be a disadvantage, especially for the Y-PCA method, where the dimensionality of $\mathscr{F}$ is typically larger.

A workaround that allows to restore the proper importance ordering of the transformed basis is to use the expression for the eigenvalues $\lambda_i$ of the correlation matrix of the map data (cf.\ eq.~(\ref{eq:eigen.app})),
\begin{equation}
\lambda_i\equiv\lambda(\bm{\mathsf{U}}_i)=\bm{\mathsf{U}}_i^T\,\mathrm{Corr}(\mathscr{D},\mathscr{D})\,\bm{\mathsf{U}}_i
\end{equation}
where $\bm{\mathsf{U}}_i$ is the corresponding (column) eigenvector. Then, if $\bm{\tilde\mathsf{U}}_i=\bm{\mathsf{R}} \bm{\mathsf{U}}_i$ is the transformed eigenvector through the same basis transformation of eq.~(\ref{eq:newbasis}), one can use $\lambda(\bm{\tilde\mathsf{U}}_i)$ as an estimate of the weight of the transformed basis vector $\bm{\epsilon}_i$ to the map data. Figure~\ref{fig:reord} shows the transformed basis of Figure~\ref{fig:bases.1D} after such reordering.

\subsection{Comparison with direct Fourier filtering}
\label{sec:FF}

Direct Fourier filtering of Stokes map frames is a popular method for tackling the problem of polarization fringe removal, because of its straightforward implementation. One must simply identify the frequency range of the fringe pattern to be removed in the spectral Fourier transform of the map frame, and suppress those frequencies before restoring the map frame by inverse Fourier transform. Fringes that are well separated in frequency space from the targeted spectral signal can easily and effectively be removed by this method. However, in general, some frequency overlap between fringes and signals is to be expected, which inevitably leads to some science data loss.

Figure~\ref{fig:AC} illustrates a typical example. The top panel on the left shows a frame from a Stokes $V$ map of the \ion{He}{1} 1083\,nm line region, with a total of 100 steps, whereas the panel right below it shows the corresponding spectral autocorrelation image. The instrument in this case was the Spectro-Polarimeter for INfrared and Optical Regions \cite[SPINOR;][]{SN06}, also deployed at the NSO/SP DST (note the spectrum is ordered by frequency rather than wavelength). The frequency interval of the fringes is well captured in Fourier space, and it is identified by the two narrow vertical strips located around $X=10$ and $X=400$, spanning the full spatial dimension. However, the map frame also contains spectral signal with a frequency range that significantly overlaps with the amplitude of the fringes in Fourier space, predominantly for positions along the spectrograph slit below $Y=100$. At those slit positions, we can expect that the reconstruction of the signal will be impacted by Fourier filtering. The panels on the right show instead the residual image of a TP-PCA reconstruction, after transformation of the PCA basis, and the exclusion of the last two fringe-carrying basis vectors. We note that the amplitude of the residual spectral signal in Fourier space has been significantly reduced, and that the separation between fringes and spectral signal is much improved, allowing for an efficient filtering of the fringes in the residual image.

Figure~\ref{fig:reconstr.1083} demonstrates in practice the effects of direct Fourier filtering in such case. The left side of the figure shows the reconstruction of the same map step in the example of Figure~\ref{fig:AC} after direct Fourier filtering of the frame. As anticipated, the spatial regions where the spectral signal has significant overlap with the fringes in frequency space suffer the most in the direct application of Fourier filtering.
The right side of the figure shows instead how the prior PCA decomposition of the Stokes map (in this case, TP-PCA) allows to attain much better results with Fourier filtering, either by selectively applying the filtering to just the (few) basis vectors visibly affected by fringes, or directly on the difference image after excluding those basis vectors from the data reconstruction (see discussion right before Sect.~\ref{sec:YPCA}). In this example, the full PCA basis of the map with 100 eigenfeatures was priorly transformed in order to confine the fringe contribution to just two basis vectors. The data reconstruction was performed using the complementary set of basis vectors, and then the difference image (corresponding to the top-right panel of Figure~\ref{fig:AC}) was Fourier filtered in order to add back the residual signal.

Figure~\ref{fig:reconstr.Ha} provides an illuminating counterexample.  In this case, we reconstructed a Stokes $Q$ frame from a map of the \ion{H}{1} H$\alpha$ line region around 656\,nm, with a total of 60 steps, also observed with the SPINOR instrument. Despite the 
complex fringe pattern in this map, direct Fourier filtering of the full frame (left) already gives very good results, comparable to what can be obtained with the selective Fourier filtering of the difference image from the TP-PCA reconstruction (right). In fact, in this case the spectral autocorrelation image of the map frame shows a clear separation between the fringe and signal contributions in frequency space, allowing a clean suppression of the fringes with minimal loss of science data.
In this case, evidently, a prior transformation of the PCA basis does not necessarily improve the fringe suppression, and in fact the right side of Figure~\ref{fig:reconstr.Ha} was obtained  by Fourier filtering the difference image after reconstructing the data without the eigenfeatures $\bm{e}_1$, $\bm{e}_2$, and $\bm{e}_6$ of the original PCA basis. Despite the effectiveness of direct Fourier filtering of the map frame in this particular example, a prior PCA of the full map, which allows to target a minimal set of PCs for the selective filtering of the fringe component, still leads to a visibly cleaner removal of the fringe signal from the map frame. This clearly demonstrates the benefit of running a PCA of a Stokes map, even in cases where the transformation of the PCA basis in order to maximally confine the fringe signal may not be necessary.

\section{Conclusions}

We studied and compared the performance of various methods for identifying and removing polarization fringes from spectro-polarimetric data based on 2-dimensional Principal Component Analysis (2D PCA) of Stokes maps. All examples were taken from data sets acquired with slit-based spectro-polarimeters, but the same methods can indistinctly be applied also to observations taken with wavelength-tunable imaging spectro-polarimeters. We analyzed in depth the performance of the \cite{TP91} algorithm of 2D PCA, which was only briefly touched upon in a previous work by \cite{Ca12}.
We compared its performance with that of the \cite{Ya04} algorithm, which was instead the focus of that previous work.

In this study, we highlighted the improvements in spectro-polarimetric data de-fringing by PCA that can be expected by subjecting the PCA basis of a Stokes map to a series of 2D rotations among its various eigenfeatures. This transformation can be set up in a quantitative and repeatable way to specifically target the fringe signal in all eigenfeatures, in order to ``sweep'' it across the basis and ``dump'' it into a minimal set of transformed basis vectors, which determine the dimensionality of the fringe subspace $\mathscr{F}$ in the data set. When perfect orthogonality between this subspace and the subspace $\mathscr{S}$ of the targeted spectral signal is realized in a particular data set, it must be expected that the basis vectors spanning the subspace $\mathscr{F}$ will be perfectly orthogonal (hence, uncorrelated) to the subspace $\mathscr{S}$.

In most practical cases, this ideal condition can never be achieved, and the problem of handling, and possibly ``rescuing'' any residual spectral signal of interest in the fringe basis vectors was also considered here.
It is found that \emph{selective} Fourier filtering of the fringe basis vectors can greatly improve the performance of 2D PCA de-fringing, as well as providing significantly better results than the more traditional \emph{direct} Fourier filtering of each individual frame of the Stokes map.

We describe one possible algorithm for the implementation of an optimal transformation of the PCA basis for the purpose of isolating the fringes into a minimal set of basis vectors (see App.~\ref{app:B}). Such algorithmic description lends itself to the development of \emph{automated} procedures for the de-fringing of spectro-polarimetric data sets.

Throughout this study, we elected to work with the \emph{correlation} matrix of the Stokes data set rather than its \emph{covariance} matrix as typically done in most PCA methods. The reason for such choice is that the creation of the covariance matrix of the data implies the subtraction of the average image from the data. In the traditional application of 2D PCA to face recognition, this average must then be added back for data reconstruction, but in the problem of Stokes data de-fringing this would inevitably imply adding back fringes to the reconstructed data. On the other hand, if the average image is left out, and this contains some significant residual spectral signal of interest, this would be inevitably lost in the reconstructed data.

The use of the covariance matrix (which was considered by \citealt{Ca12}) is thus only justified when the average image is devoid of any significant residual of the targeted spectral signal. In those cases, sometimes good data de-fringing can already be attained simply by subtracting the average image from the full data. In case of evolving fringe patterns during the Stokes map scan, the (covariance based) PCA method for de-fringing can complete the process, delivering spectro-polarimetric data sets fully usable for science.

\appendix
\section{Linear Algebra of the TP-PCA Method} \label{app:A}
We present the basic linear algebra of the \cite{TP91} approach to 2D PCA (hereafter, TP-PCA), which is needed to implement the de-fringing method for spectro-polarimetric data presented in this work.

As described in Sect.~\ref{sec:theory}, we consider a Stokes map $\mathscr{D}$ with $N$ steps,
$\{\mathscr{D}_i(x,y)\}_{i=1,\ldots,N}$, where each map step has $N_X$ wavelength points and $N_Y$ spatial points. For algebraic convenience, we reform each $N_X\times N_Y$ image $\mathscr{D}_i(x,y)$ into a $N_X N_Y$ column vector $\bm{\mathsf{D}}_i$, so that the full Stokes map can be represented by the $(N_X N_Y)\times N$ matrix $\bm{\mathsf{D}}$ with column vectors $\bm{\mathsf{D}}_i$.

In the TP-PCA approach, the $N\times N$ correlation matrix of the Stokes map is computed as
\begin{equation} \label{eq:corr.app}
\mathrm{Corr}(\mathscr{D},\mathscr{D})=\bm{\mathsf{D}}^T \bm{\mathsf{D}}\;,
\end{equation}
so its $(i,j)$ element is simply the inner product
\begin{equation}
\langle \bm{\mathsf{D}}_i,\bm{\mathsf{D}}_j \rangle\equiv \bm{\mathsf{D}}_i^T \bm{\mathsf{D}}_j\;.
\end{equation}
2D PCA involves the diagonalization of (\ref{eq:corr.app}), which is typically accomplished via the method of Singular Value Decomposition. In our case, this produces an orthogonal matrix $\bm{\mathsf{U}}$ of column eigenvectors $\bm{\mathsf{U}}_i$ with unit norm, and a corresponding, semi-positive definite, diagonal matrix $\bm{\mathsf{\Lambda}}$ of eigenvalues $\lambda_i$, with ordered diagonal elements (i.e., $\lambda_i\ge\lambda_j$ if $i>j$), such that
\begin{equation} \label{eq:eigen.app}
\bm{\mathsf{U}}^T \mathrm{Corr}(\mathscr{D},\mathscr{D})\,\bm{\mathsf{U}}=\bm{\mathsf{\Lambda}}\;.
\end{equation}

We can project the data set over the basis of eigenvectors,
\begin{equation}
\bm{\mathsf{W}}=\bm{\mathsf{D}}\,\bm{\mathsf{U}}\;,
\end{equation}
and notice that the $N$ corresponding column vectors $\bm{\mathsf{W}}_i$ form an orthogonal set, since
$\bm{\mathsf{W}}^T \bm{\mathsf{W}}=\bm{\mathsf{U}}^T\bm{\mathsf{D}}^T \bm{\mathsf{D}}\,\bm{\mathsf{U}}=\bm{\mathsf{U}}^T \mathrm{Corr}(\mathscr{D},\mathscr{D})\,\bm{\mathsf{U}}=\bm{\mathsf{\Lambda}}$. Evidently, this set is not \emph{orthonormal}. However, normalization is immediately attained by letting
\begin{equation} \label{eq:2D_basis.app}
\bm{\mathsf{E}}=\bm{\mathsf{W}}\,\bm{\mathsf{\Lambda}}^{-1/2}=\bm{\mathsf{D}}\,\bm{\mathsf{U}}\,\bm{\mathsf{\Lambda}}^{-1/2}\;.
\end{equation}
The $N$ column vectors $\bm{\mathsf{E}}_i$ of length $N_X N_Y$ can be reformed into $N$ \emph{eigenfeatures} $\bm{e}_i(x,y)$ of dimensions $N_X\times N_Y$. These eigenfeatures are orthonormal according to the inner product (\ref{eq:inner}), and therefore form a basis for the representation of the data set $\{\mathscr{D}_i(x,y)\}_{i=1,\ldots,N}$. In fact, by letting
\begin{equation}
\bm{\mathsf{D}}=\bm{\mathsf{E}}\,\bm{\mathsf{C}}\;,
\end{equation}
from eq.~(\ref{eq:2D_basis.app}) we see that the $N\times N$ matrix $\bm{\mathsf{C}}$
of expansion coefficients is uniquely determined as
\begin{equation} \label{eq:project}
\bm{\mathsf{C}}=\bm{\mathsf{E}}^T\bm{\mathsf{D}}=\bm{\mathsf{\Lambda}}^{1/2}\,\bm{\mathsf{U}}^T\;.
\end{equation}

\section{The Optimal Rotation Algorithm} \label{app:B}

We present a procedure to determine the set of optimal 2D rotations of the PCA basis necessary to confine polarization fringes to a minimal set of basis vectors. The algorithm is described for the TP-PCA method, but can easily be adapted for the Y-PCA method. 

Let $\{\mathscr{D}_i(x,y)\}_{i=1,\ldots,N}$ be a set of spectro-polarimetric images (a Stokes map), each of size $N_X\times N_Y$. We indicate with $\bar{\mathscr{D}}(x,y)$ the average image of the set. Typically, the fringes are a persistent feature of the data set, while the spectro-polarimetric signal tends to average out to some degree, especially in the case of a spatially complex target. Therefore, the average image $\bar{\mathscr{D}}(x,y)$ tends to show a weaker signal from the target with respect to the fringes. In the ideal case in which the target signal averages out completely across the Stokes map, the simple subtraction of $\bar{\mathscr{D}}(x,y)$ from the data set can be sufficient to effectively remove the fringes from each spectro-polarimetric image of the map. Otherwise, the PCA basis must be transformed in order to minimize the number of basis vectors that span the subspace $\mathscr{F}$. This can be accomplished with the following procedure:

\medskip
\begin{enumerate}
\item Select a subregion $\Sigma$ of the average image $\bar{\mathscr{D}}(x,y)$ that is as clear as possible of the target signal. This subregion will be the testing area on which the merit value for the fringe removal by rotation is going to be calculated.

\item Compute the spectral auto-correlation image of the subregion $\Sigma$. This image will show clearly identifiable, dominant bands corresponding to the range of spatial frequencies spanned by the fringes in the subregion. The 2D integral of the auto-correlation image over this interval of frequencies $\Delta k$ and the spatial dimension of the subregion $\Sigma$ can be used as the merit function of the basis rotation algorithm.

\item For each eigenfeature $\bm{e}_i(x,y)$, with $i=1,\ldots,N-1$, perform all the possible 2D rotations
\begin{eqnarray*}
\bm{\epsilon}_i(x,y)&=&\cos\theta\,\bm{e}_i(x,y)+\sin\theta\,\bm{e}_j(x,y)\;, \\
\bm{\epsilon}_j(x,y)&=&-\sin\theta\,\bm{e}_i(x,y)+\cos\theta\,\bm{e}_j(x,y)\;,
\end{eqnarray*}
for $j=i+1,\ldots,N$ and $\theta\in[-\frac{\pi}{2},+\frac{\pi}{2}]$. Compute the merit function over the subregion $\Sigma$ of $\bm{\epsilon}_i(x,y)$, and select the pair $(j,\theta)=(j_M,\theta_M)$ that produces the minimum value.

\item Perform the substitutions $\bm{e}_i(x,y)\to\bm{\epsilon}_i(x,y)$ and 
$\bm{e}_{j_M}(x,y)\to\bm{\epsilon}_{j_M}(x,y)$ in the PCA basis, and repeat steps \#3 and \#4 through the end of the loop over $i$.
\end{enumerate}
\medskip
The same algorithm can be adapted to the rotation of the 1D bases produced in the Y-PCA method. In that case, one must calculate the spectral auto-correlation function of the transformed basis vector $\bm{\epsilon}_i(x)$, and integrate it over the frequency interval $\Delta k$ of the fringes in order to compute the merit function of the transformation.

At the end of the transformation procedure, the polarization fringes will have been optimally removed from the basis vectors $\bm{\epsilon}_i$ of lower order, and confined to those of higher order.

However, it is important to clarify the meaning of ``optimally'' transformed basis in this context. According to the proposed algorithm, each eigenfeature $\bm{e}_i$ is transformed by combining with only one other eigenfeature $\bm{e}_j$ of the PCA basis, where $j>i$. The algorithm determines the optimal pairing of eigenfeatures, but generally none of those available in the PCA basis will lead to a perfect cancellation of the unwanted signal in the transformed basis vector $\bm{\epsilon}_i$. However, when the rotation procedure is completed, the transformed basis $\{\bm{\epsilon}_i\}_{i=1,\ldots,N}$ will contain a  new set of basis vector, which may be ``more optimal'' than the original PCA basis for the purpose of canceling the (residual) unwanted signal from each eigenfeature.\footnote{By construction, this is indeed the case, since the algorithm naturally tends to group like signals into a minimal set of transformed basis vectors.} In other words, it is possible to improve the realization of the orthogonality between $\mathscr{F}$ and $\mathscr{S}$ by iterating the rotation algorithm over the transformed basis 
$\{\bm{\epsilon}_i\}_{i=1,\ldots,N}$.


\begin{thebibliography}{}
%
\bibitem[\protect\citeauthoryear{Birkhoff \& Mac Lane}{1953}]{BM53}
Birkhoff, G., \& Mac Lane, S.~1953, A Survey of Modern Algebra (MacMillan: New York)
%
\bibitem[\protect\citeauthoryear{Cao et al.}{2010}]{Ca10}
Cao, W., Gorceix, N., Coulter, R., Ahn, K., Rimmele, T. R., \& Goode, P. R.~2010, AN, 331, 636
%
\bibitem[\protect\citeauthoryear{Casini \& Landi Degl'Innocenti}{2008}]{CL08}
Casini, R., \& Landi Degl'Innocenti, E.~2008, Astrophysical Plasmas, in
Plasma Polarization Spectroscopy, eds.\ T.~Fujimoto \& A.~Iwamae
(Springer: Berlin), 247
%
\bibitem[\protect\citeauthoryear{Casini, Judge, \& Schad}{2012}]{Ca12}
Casini, R, Judge, P.~G., \& Schad, T.~A.~2012, \apj, 756, 193
%
\bibitem[\protect\citeauthoryear{Cavallini}{2006}]{Ca06}
Cavallini, F.~2006, \solphys, 236, 415
%
\bibitem[\protect\citeauthoryear{Clarke}{2004}]{Cl04}
Clarke, D.~2004, J.\ Opt.~A, 6, 1036
%
\bibitem[\protect\citeauthoryear{Collados et al.}{2013}]{Co13}
Collados, M., Bettonvil, F., Cavaller, L., Ermolli, I., Gelly, B., P\'erez, A., Socas-Navarro, H., Soltau, D., Volkmer, R., \& EST Team~2013, Mem. SAIt, 84, 379
%
\bibitem[\protect\citeauthoryear{Harrington et al.}{2017}]{Ha17}
Harrington, D. M., Snik, F., Keller, C. U., Sueoka, S. R., \& van Harten, G.~2010, JATIS, 3, 048001 (DOI:10.1117/1.JATIS.3.4.048001)
%
\bibitem[\protect\citeauthoryear{Hasan et al.}{2010}]{Ha10}
Hasan, S. S., Soltau, D., K\"archer, H., S\"u\ss, M.,\& Berkefeld, T.~2010, AN, 331, 628
%
\bibitem[\protect\citeauthoryear{Jaeggli et al.}{2010}]{Ja10}
Jaeggli, S.~A., Lin, H., Mickey, D.~L., Kuhn, J.~R., Hegwer, S.~L.,
Rimmele, T.~R., and Penn, M.~J.~2010, Mem.\ SAIt, 81, 763
%
\bibitem[\protect\citeauthoryear{Jolliffe}{2002}]{Jo02}
Jolliffe, I.~T.~2002, Principal Component Analysis, 2nd ed.\ (Springer:
New York)
%
\bibitem[\protect\citeauthoryear{Landi Degl'Innocenti \& Landolfi}{2004}]{LL04}
Landi Degl'Innocenti, E., \& Landolfi, M.~2004, Polarization in Spectral
Lines (Springer: Dordrecht)
%
\bibitem[\protect\citeauthoryear{Lites}{1991}]{Li91}
Lites, B.~W.~1991, Fringes in Polarizing Optical Elements, in Solar Polarimetry, Proc. 11th Sacramento Peak Summer Workshop, ed. L.~J.\ November (Sunspot: NSO), 173
%
\bibitem[\protect\citeauthoryear{Liu, Deng, \& Ji}{2014}]{Li14}
Liu, Z., Deng, Y., \& Ji, H.~2014, The Chinese Giant Solar Telescope, in Nature of Prominences and their Role in Space Weather, eds.\ B.~Schmieder, J.-M.~Malherbe, \&  S.~T.~Wu, Proc. IAU, 300, 349
%
\bibitem[\protect\citeauthoryear{Pearson}{1901}]{Pe01}
Pearson, K.~1901, Phil.\ Mag., 2, 559
%
\bibitem[\protect\citeauthoryear{Rojo \& Harrington}{2006}]{Ro06}
Rojo, P.~M., \& Harrington, J.~2006, \apj, 649, 553
%
\bibitem[\protect\citeauthoryear{Semel}{2003}]{Se03}
Semel, M.~2003, A\&A, 401, 1
%
\bibitem[\protect\citeauthoryear{Skumanich \& L\'opez Ariste}{2002}]{SL02}
Skumanich, A.~P., \& L\'opez Ariste, A.~2002, \apj, 570, 379
%
\bibitem[\protect\citeauthoryear{Socas-Navarro et al.}{2006}]{SN06}
Socas-Navarro, H., Elmore, D., Pietarila, A., Darnell, A., Lites, B.~W., Tomczyk, S., \& Hegwer, S.~2006, \solphys, 235, 55
%
\bibitem[\protect\citeauthoryear{Stenflo}{1994}]{St94}
Stenflo, J.~O.~1994, Solar Magnetic Fields (Kluwer: Dordrecht)
%
\bibitem[\protect\citeauthoryear{Tritschler et al.}{2016}]{Tr16}
Tritschler, A., Rimmele, T. R., Berukoff, S., Casini, R., Kuhn, J. R., Lin, H., Rast, M. P., McMullin, J. P., Schmidt, W., W\"oger, F., \& DKIST Team~2016, AN, 337, 1064
%
\bibitem[\protect\citeauthoryear{Trujillo Bueno}{2010}]{TB10}
Trujillo Bueno, J.~2010, Recent Advances in Chromospheric and Coronal
Polarization Diagnostics, in Magnetic Coupling between the Interior 
and Atmosphere of the Sun, Proc.\ Astrophysics and Space Science, 
eds.\ S.~S.\ Hasan \& R.~J.\ Rutten (Springer: Berlin), 118
%
\bibitem[\protect\citeauthoryear{Turk \& Pentland}{1991}]{TP91}
Turk, M.~A., \& Pentland, A.~P.~1991, JoCN, 3, 71
%
\bibitem[\protect\citeauthoryear{Yang et al.}{2004}]{Ya04}
Yang, J., Zhang, D., Frangi, A.~F., \& Yang, J.-Y.~2004, IEEE Trans.\ PAMI,
26, 131

\end{thebibliography}
\end{document}